\documentclass[runningheads]{llncs}
\usepackage[T1]{fontenc}
\usepackage{graphicx}
\usepackage{booktabs}
\usepackage[misc]{ifsym}
\newcommand{\corr}{(\Letter)}
\newtheorem{observation}{Observation}

\usepackage{amsmath,amssymb,amsfonts}
\usepackage{algorithm}
\usepackage{algorithmic}
\usepackage{multicol}
\usepackage{multirow}
\usepackage{subfigure}

\begin{document}

\title{A Hierarchical Scale-free Graph Generator under Limited Resources}

\titlerunning{Hierarchical Scale-free Generator under Limited Resources}

\author{Xiaorui Qi \and
Yanlong Wen\corr \and
Xiaojie Yuan}

\authorrunning{X. Qi et al.}

\institute{College of Computer Science, Nankai University, China \email{qixiaorui@mail.nankai.edu.cn, \{wenyl, yuanxj\}@nankai.edu.cn}}

\maketitle              

\begin{abstract}
Graph generation is one of the most challenging tasks in recent years, and its core is to learn the ground truth distribution hiding in the training data.
However, training data may not be available due to security concerns or unaffordable costs, which severely blows the learning models, especially the deep generative models.
The dilemma leads us to rethink non-learned generation methods based on graph invariant features.
Based on the observation of scale-free property, we propose a hierarchical scale-free graph generation algorithm.
Specifically, we design a two-stage generation strategy.
In the first stage, we sample multiple anchor nodes to further guide the formation of substructures, splitting the initial node set into multiple ones.
Next, we progressively generate edges by sampling nodes through a degree mixing distribution, adjusting the tolerance towards exotic structures via two thresholds.
We provide theoretical guarantees for hierarchical generation and verify the effectiveness of our method under 12 datasets of three categories.
Experimental results show that our method fits the ground truth distribution better than various generation strategies and other distribution observations.

\end{abstract}

\section{Introduction}
Graph generation task, widely used for scenarios like social analysis~\cite{ref:social2021,ref:social2022,ref:social2024}, molecule generation~\cite{ref:bio2019,ref:bio2023,ref:bio2024}, and structure modeling~\cite{ref:ss2022,ref:ss2023,ref:ss2024}, is one of the hottest researches in recent years.
Maintaining the distribution similarity between the generated and ground truth graphs is the core and one of the challenges of this task.
Deep Generative Models (DGM) are the mainstream solutions nowadays~\cite{ref:DGM,ref:CF,ref:GraphRNN,ref:GraphARM,ref:GraphILE}.
DGM learns the latent distribution features hiding in the training data, generating graphs closer to the requirements.

\begin{figure}[!t]
    \centering
    \includegraphics[width=0.8\linewidth]{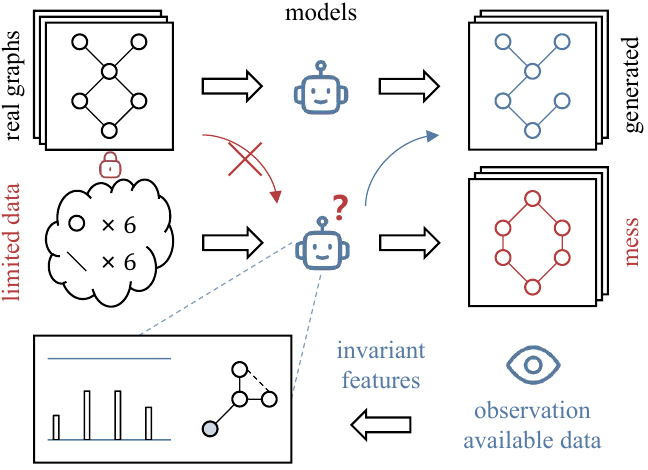}
    \caption{An illustration of deep learning models with limited/no training data failing to generate graphs due to safety propose or high price (in red). However, with invariant features dragging from observation on universal data available, we can generate graphs fitting ground truth distribution well without learning (in blue).}
    \label{fig:pre}
\end{figure}

However, training data may not be available due to security concerns or high data acquisition costs~\cite{ref:trans2024}, striking learning models severely, especially DGM.
Figure~\ref{fig:pre} illustrates this limitation.
Only plain graph property data, such as the number of nodes and edges, are available when we limit the training data.
The learnable distribution space of DGM switches from the training set to the search space of node permutations, resulting in the generation of mess graphs out of distribution (in red).
Some studies have proposed solutions to this problem, such as graph transfer learning~\cite{ref:trans2019,ref:trans2023,ref:trans2024}, which realizes graph generation in restricted scenarios by learning invariant features in other training data that can be easily obtained.
Nevertheless, the quality of such solutions depends heavily on the quality of the training data before transfer.
There is heterogeneity in the data distribution of different scenarios~\cite{ref:DS}.
The model performance is unsatisfactory if there are considerable differences in the invariant features.
For example, social networks have many clusters, while molecules rely on functional groups with unique semantics.
Models can not learn sufficient invariant features transferring one scene above to another.
Improving the quality of training data further increases the acquisition cost, returning to the previous limitations.

This dilemma urges us to rethink non-learned generation methods based on graph invariant features.
Traditional methods have been designed on universal observations maintained in real-world scenarios, such as randomness~\cite{ref:ER}, small world~\cite{ref:WS}, scale-free~\cite{ref:BA,ref:CL,ref:RMAT}, etc.
These observations are equal to invariant features independent of the training data, satisfying the properties of graphs under most scenarios.
Among them, we pay attention to the scale-free property of graphs.
Scale-free, also known as power-law, heavy-tailed, etc., refers to a phenomenon in which a small number of nodes in networks have a large number of connections, while most nodes have only a few.
Many systems in the real world have this property~\cite{ref:survey2022,ref:BA}.
It will serve as a significant invariant feature that effectively organizes the basic graph information to direct the generation.

But, instead of DGM learning the real-world distribution accurately, a selected invariant feature often orients to a class of graph sets with its hand-crafted characteristics.
For example, Chung-Lu (CL)~\cite{ref:CL} generates strictly scale-free networks and does not perform well in generating other graphs that do not follow this degree distribution.
Some more methods~\cite{ref:BTER,ref:Darwini,ref:MMSB} serve the corresponding set of graphs depending on the user-defined feature constraints, while others~\cite{ref:gMark,ref:LinkBench,ref:LinkGen} define particular scenarios, leaving the decision to randomness. 
When learning approaches are limited as a resource, the unapproachable power of predefined features kills the generalization ability of non-learned methods~\cite{ref:survey2022}.
Therefore, whether enhancing the generality of invariant features or weakening selected features by noises, restrictions, etc., can improve the support of non-learned methods for rich graph categories.

Overall, to devise graph generation methods under limited resources scenarios, we have the following challenges:
\textbf{(1) Capture powerful invariant features.}
The ground truth distribution cannot be learned directly due to the difficulty of obtaining training data.
We need a generation strategy based on a powerful invariant feature covering the majority of scenes.
\textbf{(2) Weaken the influence of manual design.}
Traditional methods are engineered for a family of graphs with selected features, leading to poor scalability~\cite{ref:survey2022}.
We need to weaken the influence of particular features to support the generation of diverse structures.
\textbf{(3) Generate graphs that fit the ground truth distribution.}
The effect of invariant feature selection will eventually be reflected in the distance evaluation between the generated and real-world graphs.
We need to generate graphs closer to the ground truth.

To solve the above problems, we propose a hierarchical scale-free graph generation algorithm, which generates graphs under the observation of the scale-free property.
Specifically, we design a two-stage generation strategy.
In the first stage, we sample a set of anchor nodes to further guide the formation of substructures, splitting the node set into multiple ones.
It corresponds to challenge 1, where we follow the scale-free invariant features.
In the second stage, we employ preferential attachment to weight the scale-free distribution to alleviate the feature limitations mentioned in challenge 2.
We progressively generate the remaining edges by sampling node pairs through a degree mixing distribution.
Finally, facing challenge 3, we leverage two basic graph information as thresholds to adjust the algorithm tolerance for generating exotic structures and further enrich the search space well-fitted to ground truth distribution.
We provide theoretical underpinning that guarantees that the generative process is interpretable.

In summary, our contributions are as follows:
\begin{itemize}
    \item We explore the possibility of generating graphs in the setting of limited resources and rethink the generation strategy only through basic graph properties.
    \item We design a hierarchical scale-free graph generator based on the scale-free properties under limited resources, which leverages degree mixing to weaken the restriction and creates graphs closer to the real-world distribution.
    \item Experiments on 12 datasets in three categories give a thorough perspective to evaluate our work. Results show that our method can cover most scenarios when the ground truth distribution is unknown.
\end{itemize}

\section{Related Work}
\subsection{Traditional Methods}
In the early stage, traditional graph generation methods rely on human-made characteristics or particular strict constraints.
Erd\H{o}s-R\'enyi (ER)~\cite{ref:ER} and Watts-Strogatz (WS)~\cite{ref:WS} are the most famous graph generation paradigms.
The former is $\mathcal{G}_{np}$, where $n$ is the number of nodes and $p$ is the connection probability between every node pair.
The latter is $\mathcal{G}(n, k, p)$, which requires that each node associates with its $k$ neighbors at initialization.
Then, one of the endpoints of each edge will change with probability $p$.
Barab\'asi-Albert (BA)~\cite{ref:BA} pays attention to the preferential attachment phenomenon, defines $p$ as the degree proportion where $p_u = \frac{d_u}{\sum_{v \in \mathcal{N}(u)} d_v}$, and proposes a progressive generation method that gradually adds nodes.
Therefore, BTER~\cite{ref:BTER} created a hierarchical generation method combining ER and BA, leveraging the node proportion to connect multiple ER pre-generated graphs.

Similar to hierarchy, another branch of recursive generation exists.
These methods generate graphs from an initial graph $\mathcal{G}_0$.
R-MAT~\cite{ref:RMAT} and Kronecker~\cite{ref:Kronecker} notice the adjacency matrix, creating self-similarity graphs by copying the initial one.
Meanwhile, Mixed-Membership Stochastic Block (MMSB)~\cite{ref:MMSB} shares the same idea of parsing graphs into multiple blocks as BTER, which builds a reliable community analysis.
\subsection{Deep Generative Models}
Considering the limitations of traditional methods, researchers have shifted their attention to Deep Generative Models (DGM) in recent years.
Depending on the technique employed, DGM can be divided into four categories: Variational Auto-Encoders (VAE)~\cite{ref:VAE}, Generative Adversarial Networks (GAN)~\cite{ref:GAN}, Normalizing Flows (NF)~\cite{ref:NF}, and Diffusion models~\cite{ref:Diff}.
VAE and NF transform the original graph feature space into a low-dimensional vector, where the former leverages a well-trained Graph Neural Network (GNN) based encoder and decoder, and the latter uses a sequence of invertible functions.
GAN implicitly learns the distribution of the original graph by training a pair of generators and discriminators, which play against each other.
Denoising diffusion models originated from the field of computer vision.
The core idea is to perturb the original distribution through a designed noise model and then train a learnable denoising process to recover the original graph data from the noise.

We use an earlier contemporaneous GraphRNN~\cite{ref:GraphRNN}, which leverages recurrent neural networks as the backbone, to represent the single class of technical schemes.
Meanwhile, the majority of popular DGM uses a mixed architecture.
For example, GraphARM~\cite{ref:GraphARM} relies on an autoregressive diffusion scheme to improve the generation speed of a single solution.
GraphILE~\cite{ref:GraphILE} designs the pipeline based on graph coarsening and expansion techniques and integrates the denoising process for training.
CatFlow~\cite{ref:CF} introduces variational inference to flow matching, amplifying the advantages of both.
For more details on the graph generation tasks, please refer to~\cite{ref:survey2020,ref:survey2022,ref:survey2023}.

\section{Preliminary}
\begin{figure*}[!t]
    \centering
    \includegraphics[width=\linewidth]{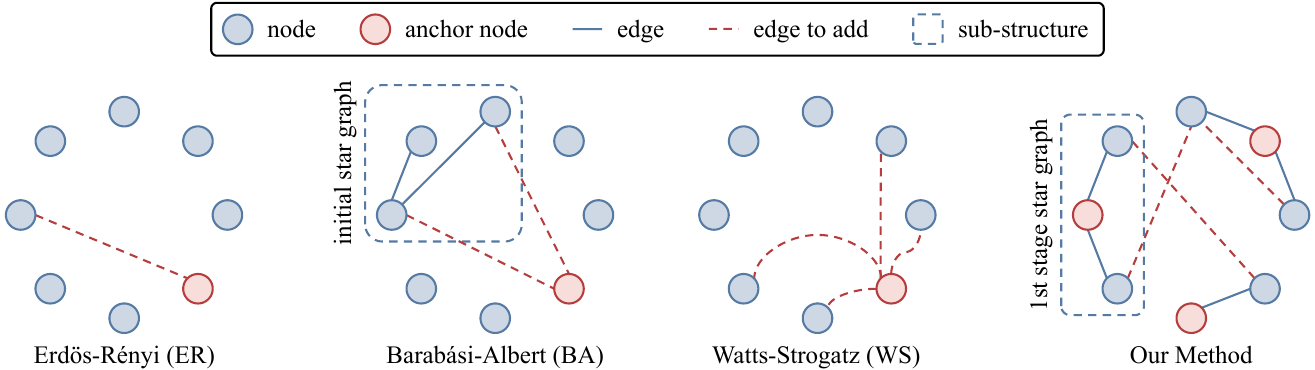}
    \caption{Generation strategy comparison between several baselines and our method.}
    \label{fig:framework}
\end{figure*}
\begin{definition}[Graph]
    A graph can form as $\mathcal{G}=\{\mathcal{V}, \mathcal{E}\}$, where $\mathcal{V}$ and $\mathcal{E}$ represent nodes and edges set, and $N$ and $M$ are the number of nodes the edges, respectively.
    $A[u, v] \in \{0, 1\}^{N \times N}, u, v \in \mathcal{V}$ is the adjacency matrix, where $A[u, v] = 1$ shows there is an edge between node $u$ and node $v$.
    We use $d_u$ to represent the degree of node $u$.
    For an undirected graph, $d_u = \sum_{v \in \mathcal{V}} A[u, v]$.
    We use $(N, M)$ as a shorthand for a graph $\mathcal{G}$ with $N$ nodes and $M$ edges.
    \label{def:graph}
\end{definition}

Given a set of graphs $\mathbb{G} = \{\mathcal{G}_1, ..., \mathcal{G}_{|\mathbb{G}|}\}$ sampled from a particular scenario $\mathcal{S}$, each graph is under the distribution determined by the characteristics of $\mathcal{S}$, where $\mathcal{G} \sim p_{\mathcal{S}}(\mathcal{G})$.
Graph generation is essentially a task of sampling from the given distribution $p_{\mathcal{S}}$.
However, unfortunately, $p_{\mathcal{S}}$ is not always known.
Sometimes, even $\mathbb{G}$ is unknown due to security concerns or unaffordable costs.
In this case, we can only obtain basic graph information like the number of nodes and edges and generate random graphs via the universal observation $p_{\mathcal{O}}$.
Our method builds on existing observations as follows:
\begin{observation}
    For real-world systems, most of them show the heavy-tailed degree distribution~\cite{ref:survey2022}.
    In other words, the degree distribution of these graphs has the power-law or scale-free characteristics.
    \label{ob:main}
\end{observation}

The scale-free property of graphs is not a new concept, and researchers have studied it extensively~\cite{ref:BA}.
A general conclusion states that the degree distribution of random graphs is close to the Poisson distribution, which is
\begin{equation}
    P(X = d) = e^{-\lambda} \frac{\lambda^{d}}{d!},
    \label{eq:poisson}
\end{equation}
where $d$ and $\lambda$ are the selection and expectation of the node degree, respectively.
We use $P(k_0|\lambda_0)$ as a shorthand for $P(X = k_0)$ with $\lambda=\lambda_0$.
Notations and definitions are in Appendix A.
We also provide a further explanation of motivation in Appendix B.
\begin{definition}[Problem Statement]
    Given a set of graphs $\mathbb{G}$, we aim to create a generated set $\hat{\mathbb{G}}$ from the restricted information (e.g., $N$, $M$) under the premise that the ground truth distribution $p_{\mathcal{S}}$ is unknown.
    Since $p_{\mathcal{S}}$ is limited, the goal is to leverage some universal observations $p_{\mathcal{O}}$ (e.g., Observation~\ref{ob:main}) to make $\hat{\mathbb{G}}$ closer to $\mathbb{G}$.
    \label{def:problem}
\end{definition}

\section{Methodology}
\subsection{Overview}
Figure~\ref{fig:framework} shows an overview of our method, comparing the generation strategy with several baselines.
Our method generates graphs and updates the node selection probabilities hierarchically.
In the first stage, we parse the original nodes to build initial star graphs led by anchor nodes (Sec. IV-B). 
In the second stage, we prioritize ensuring the connectivity between previous subgraphs.
Next, we design the node selection strategy to generate the remaining edges through a degree mixing distribution (Sec. IV-C). 
The article followed is structured in order of the graph generation.
\subsection{First Step of Generation}
Generating a graph $\mathcal{G}$ directly from $N$ and $M$ is equivalent to randomly sampling from the search space of $\binom{\binom{N}{2}}{M}$, which is a bit low probability.
Inspired by the recursive~\cite{ref:RMAT,ref:MMSB,ref:Kronecker} and multi-level~\cite{ref:BTER,ref:Darwini} generation methods, we design our work in a hierarchical stepwise scheme to improve the rationality of the distribution of the graph generated.
In the first generation step, we extend the initial graph like~\cite{ref:BA} into multiple substructures, defined as follows.
\begin{definition}[Substructure]
    A substructure $\mathcal{G}_{sub}^{(i)}$ is a smaller graph derived from the original graph $\mathcal{G}$, where $\mathcal{V}_{sub}^{(i)} \subseteq \mathcal{V}$ and is set of nodes.
    $(N_{sub}^{(i)} + 1, N_{sub}^{(i)})$ is a star graph that exists a node $v_0^{(i)}$ connected to all other nodes $v_j^{(i)} (j = 1, ..., N_{sub}^{(i)})$.
    We call $v_0^{(i)}$ an anchor node of $\mathcal{G}_{sub}^{(i)}$.
    \label{def:sub}
\end{definition}

Our method builds a parsing process from a given graph $(N, M)$ to a substructure sequence $\mathcal{G}_{sub} = \{\mathcal{G}_{sub}^{(0)}, ..., \mathcal{G}_{sub}^{(l-1)}\}$, where $\mathcal{G}_{sub}^{(i)} = (d_i + 1, d_i)$, $d_i$ is the node degree of $v_0^{(i)}$, and $l$ is the number of substructures.
We sample the degree sequence $D = \{d_0, ..., d_{l-1}\}$ of anchor nodes through Equation~\ref{eq:poisson}, guiding the generation of $\mathcal{G}_{sub}$, where $D \sim P(\lambda), \sum_{i=0}^{l-1} d_i = N - l$.
Given $(N, M)$, we leverage the available information to estimate the average degree expectation, where $\lambda = \frac{2M}{N} = \bar{D}$.

When sampling $D$, it cannot avoid the appearance of exotic structures (e.g., star graphs), though the probability of large degrees is too low to zero.
Therefore, we focus on two parameters to control the tolerance for unique structures when generating graphs.
Similar to~\cite{ref:BA,ref:WS}, the first parameter we take into account is the max degree $d_{max}$.
The 1-hop neighborhood of a center node shows wealthy local information, which builds the aggregation in the message-passing mechanism~\cite{ref:GCN}.
Inspired by this, we limit the number of edges elicited by each node with the $d_{max}$ to fit the ground truth situation.
And that is why we sample anchors first and build the star graph instead of any other structures.
The second parameter is a truncation against Equation~\ref{eq:poisson}.
For large degrees much greater than the average, denoted by $d_\infty \gg \bar{D}$, it shows that 
\begin{equation}
    P(d_\infty | \bar{D}) \simeq 0 < P(0 | \bar{D}).
    \label{eq:pk}
\end{equation}
Since a connected graph does not have isolated nodes of degree 0 and node with extremely high degrees are rare, we can approximately remove $d_\infty$ or more than $d_\infty$ cases.
We define $k$ to be the smallest positive integer satisfying this equation.
\subsection{Hierarchical Scale-free Generation Strategy}
\begin{figure}[!t]
    \centering
    \includegraphics[width=\linewidth]{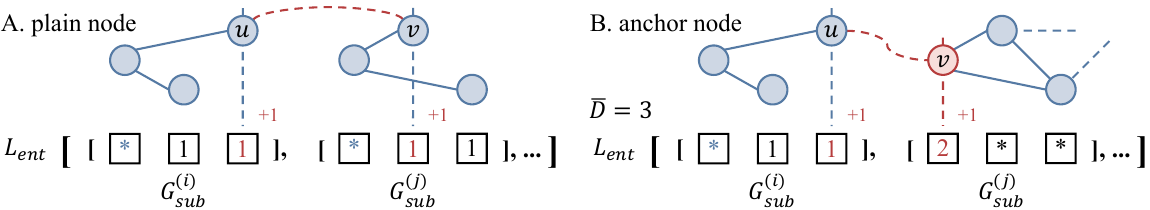}
    \caption{Illustration of the edge generation. (A) connection to plain nodes. (B) connection to anchor nodes.}
    \label{fig:2stage}
\end{figure}
After the first stage, we split the whole graph into a sequence of substructures, ensuring the generation of $N$ nodes and carrying a small amount of edges.
In the second stage, we generate the remaining edges gradually.
Instead of~\cite{ref:ER} directly specifying the probability of edge generation, we build edges by selecting two endpoints.
Inspired by~\cite{ref:BTER,ref:MMSB}, we expand $D$ into a list of entries $L_{ent}$ as follows.
\begin{definition}[Entry]
    $L_{ent}$ is a list of entries, and $L_{ent}[i]$ represents the degree sequence $D_{sub}^{(i)}$ of $\mathcal{G}_{sub}^{(i)}$, where $D_{sub}^{(i)}[0]$ is the degree of the $v_0^{(i)}$.
    We use $L_{ent}[i][j]$ to refer to the node $v_j^{(i)}$ located in $\mathcal{G}_{sub}^{(i)}$, forming an entry.
    For example, in Figure~\ref{fig:2stage}, $L_{ent}[0][2]$ represents node $u$.
    \label{def:entry}
\end{definition}

This auxiliary structure costs additional $\mathcal{O}(N)$ space, assisting the probability list generation.
We independently sample nodes $u$ and $v$ to form the edge $e_{uv}$.
For each node $v_j^{(i)}$, the probability is as follows:
\begin{equation}
    Pr(v_j^{(i)}) = s_{sub}^{(i)} \cdot P(L_{ent}[i][j] + 1|\bar{D}).
    \label{eq:pro}
\end{equation}
The probability consists of two parts: $s_{sub}^{(i)}$ represents the probability of entering $\mathcal{G}_{sub}^{(i)}$, and $P(L_{ent}[i][j] + 1|\bar{D})$ represents the sampling probability of the node jumping to the next degree of the current one based on Equation~\ref{eq:poisson}.
$s_{sub}^{(i)}$ refers to the preferential attachment theory proposed by~\cite{ref:BA}.
\begin{theorem}[Preferential Attachment]
    The more connections between nodes, the more likely it is to receive new connections.
    Consequently, there are central nodes or hubs in networks linking to abundant new nodes.
    \label{th:ss}
\end{theorem}

Since the first stage creates independent substructures, we naturally use $N_{sub}^{(i)}$ to calculate the probability of connecting $\mathcal{G}_{sub}^{(i)}$ with the formula:
\begin{equation}
    s_{sub}^{(i)} = \frac{N_{sub}^{(i)}}{N}.
    \label{eq:ss}
\end{equation}
Each node located in $\mathcal{G}_{sub}^{(i)}$ shares Equation~\ref{eq:ss} and adjusts itself according to the next-degree distribution of the current one, which is $P(L_{ent}[i][j] + 1|\bar{D})$.
This design is a variant of~\cite{ref:CL} based on the desired degree theory.
\begin{theorem}[Desired Degree]
    It can estimate the probability of selecting nodes via the occurrence of nodes with a particular degree $d_*$, called the desired degree.
\end{theorem}

Our method samples nodes according to the current $L_{ent}$ and updates constantly.
According to the assumptions in~\cite{ref:CL,ref:BTER}, the process of selecting nodes $u$ and $v$ is independent.
Thus, the probability of an edge is
\begin{equation}
    P_{e_{uv}} = \bar{Pr}(u) \cdot \bar{Pr}(v),
    \label{eq:PE}
\end{equation}
where $\bar{Pr}(\cdot)$ is the probability after normalization, satisfying the probability sum of 1, that is
\begin{equation}
    \sum_{i=0}^{l-1}\sum_{j=0}^{N_{sub}^{(i)}}\bar{Pr}(v_j^{(i)})=1.
    \label{eq:sum}
\end{equation}

In addition, our method prioritizes establishing connections between non-anchor nodes.
Figure~\ref{fig:2stage} illustrates the edge generation strategy under different node types.
Given the max degree $d_{max}$ constraint, we set $Pr(v_0^{(i)})=0$ if existing non-anchor nodes whose degrees are less than $d_{max}$.
We use $*$ to mask the corresponding positions $L_{ent}[i][0]$, as shown in Figure~\ref{fig:2stage}(A).
After sampling the nodes $u$ and $v$, the change of degree updates in $L_{ent}$ from 1 to 2.
As the number of edges increases, as shown in Figure~\ref{fig:2stage}(B), there will be a case where all non-anchor nodes have connected $d_{max}$ edges like $\mathcal{G}_{sub}^{(j)}$.
In this case, we mask the $L_{ent}$ of the non-anchor nodes with $*$ and restore the anchor node to its current degree.
For example, in Figure~\ref{fig:2stage}(B), we set $L_{ent}[j][0] = 2$.
After sampling the node pair, the update is the same as in 3(A). 
If the anchor node $v_0^{(i)}$ exceeds $d_{max}$, too, we mask this $\mathcal{G}_{sub}^{(i)}$ without visiting anymore.
\subsection{Discussion}
We discuss some issues further in Appendix C.
The pseudo-code is in C.1, the theoretical proof is in C.2, and the time complexity and limitation of our work are in C.3 and C.4, respectively.
We further evaluate the time complexity of our method in experiments with some large datasets, please see Appendix E.4 for details.
\section{Experiments}
\subsection{Experimental Settings}
\subsubsection{Datasets}
We perform experiments on eight real-world from two categories and four synthetic datasets to evaluate our work thoroughly.
More details about datasets and implementations are in Appendix D.
\subsubsection{Baselines}
We compare five traditional baselines: Erd\H{o}s-R\'enyi (ER)~\cite{ref:ER}, Barab\'asi-Albert (BA)~\cite{ref:BA}, Watts-Strogatz (WS)~\cite{ref:WS}, Mixed-Membership Stochastic Block (MMSB)~\cite{ref:MMSB}, and Kronecker~\cite{ref:Kronecker}.
We also pick several popular and recent models: GraphRNN~\cite{ref:GraphRNN}, GraphARM~\cite{ref:GraphARM}, and GraphILE~\cite{ref:GraphILE}.
\subsubsection{Evaluation Metrics}
We use the Maximum Mean Discrepancy (MMD)~\cite{ref:MMD} to evaluate the quality of generated graphs.
Following~\cite{ref:GraphRNN}, we compute MMD scores for the average degree, clustering coefficient, and orbit count.
\subsection{Main Results}
\begin{table*}[!t]
    \centering
    \caption{MMD evaluation between baselines and our method.}
    \begin{tabular}{c|ccc|ccc|ccc}
        \hline
        \multirow{2}{*}{} & \multicolumn{3}{c}{ENZYMES} & \multicolumn{3}{c}{deezer\_ego\_nets} & \multicolumn{3}{c}{CLUS} \\
        Methods & deg. & clus. & orbit & deg. & clus. & orbit & deg. & clus. & orbit \\
        \hline
        ER & \underline{0.2488} & 0.6522 & 0.3775 & 0.2865 & 0.6186 & 0.3814 & 0.1557 & 0.3700 & 0.1564 \\
        BA & 0.8302 & \textbf{0.3198} & 0.7207 & \underline{0.2543} & \underline{0.3494} & \underline{0.2149} & \underline{0.0880} & \underline{0.2100} & \underline{0.0606} \\        
        WS & 0.6769 & 0.9314 & \textbf{0.0657} & 0.4242 & 0.7007 & 0.5348 & 0.3510 & 0.6344 & 0.2894 \\
        MMSB & 1.6808 & 1.3862 & 0.3420 & 0.7714 & 1.2728 & 1.0237 & 0.8183 & 1.0930 & 0.5017 \\
        Kronecker & 1.4012 & 1.3862 & 0.6402 & 0.4082 & 1.2285 & 0.4832 & 0.9960 & 1.3256 & 0.7888 \\
        \hline
        Ours & \textbf{0.1811} & \underline{0.4621} & \underline{0.1271} & \textbf{0.0371} & \textbf{0.0668} & \textbf{0.0026} & \textbf{0.0840} & \textbf{0.1911} & \textbf{0.0509} \\
        \hline
        \hline
        \multirow{2}{*}{} & \multicolumn{3}{c}{MUTAG} & \multicolumn{3}{c}{IMDB-BINARY} & \multicolumn{3}{c}{EGO} \\
        Methods & deg. & clus. & orbit & deg. & clus. & orbit & deg. & clus. & orbit \\
        \hline
        ER & 0.1688 & 0.2501 & 0.0119 & \underline{0.0812} & 0.7398 & \underline{0.1525} & 0.1425 & 0.2875 & 0.0147 \\
        BA & 1.0271 & 1.3066 & 0.6054 & 0.1614 & 0.9386 & 0.1583 & \underline{0.0413} & 0.0000 & \underline{0.0087} \\
        WS & \underline{0.0821} & \textbf{0.0233} & \underline{0.0095} & 0.2443 & 0.8787 & 0.4040 & - & - & - \\
        MMSB & 0.9803 & 1.3099 & 0.0539 & 0.3857 & \underline{0.7217} & 0.4481 & - & - & - \\
        Kronecker & 1.7907 & 1.8785 & 0.0603 & 1.1343 & 1.8764 & 0.7516 & 1.4997 & 0.0000 & 0.4574 \\
        \hline
        Ours & \textbf{0.0355} & \underline{0.0561} & \textbf{0.0014} & \textbf{0.0393} & \textbf{0.7161} & \textbf{0.1436} & \textbf{0.0382} & \textbf{0.0000} & \textbf{0.0080} \\
        \hline
        \hline
        \multirow{2}{*}{} & \multicolumn{3}{c}{NCI1} & \multicolumn{3}{c}{IMDB-MULTI} & \multicolumn{3}{c}{GRID} \\
        Methods & deg. & clus. & orbit & deg. & clus. & orbit & deg. & clus. & orbit \\
        \hline
        ER & 0.2125 & 0.2014 & 0.0242 & \underline{0.0317} & \underline{0.4710} & \underline{0.0663} & \underline{0.3799} & 0.5371 & \underline{0.0792} \\
        BA & 1.0987 & 1.1966 & 0.9007 & 0.2825 & 1.0538 & 0.2122 & 1.1267 & 1.4074 & 0.7096 \\
        WS & \underline{0.1553} & \underline{0.0278} & \underline{0.0063} & 0.0761 & 0.4913 & 0.1314 & 1.0814 & \textbf{0.0470} & 0.3745 \\
        MMSB & 0.8707 & 0.2793 & 0.0434 & 0.3850 & 0.6088 & 0.1742 & 1.6535 & 1.6784 & 0.3293 \\
        Kronecker & 1.6345 & 1.6594 & 0.0414 & 1.7928 & 1.9521 & 0.6108 & 1.9735 & 1.7947 & 0.4274 \\
        \hline
        Ours & \textbf{0.0340} & \textbf{0.0263} & \textbf{0.0002} & \textbf{0.0160} & \textbf{0.3882} & \textbf{0.0416} & \textbf{0.2459} & \underline{0.4081} & \textbf{0.0148} \\
        \hline
        \hline
        \multirow{2}{*}{} & \multicolumn{3}{c}{PROTEINS} & \multicolumn{3}{c}{REDDIT-BINARY} & \multicolumn{3}{c}{TREE} \\
        Methods & deg. & clus. & orbit & deg. & clus. & orbit & deg. & clus. & orbit \\
        \hline
        ER & \underline{0.2927} & 0.5814 & 0.4779 & \underline{1.0232} & \underline{0.0908} & 1.0088 & \underline{0.0841} & 0.2227 & \underline{0.0020} \\
        BA & 0.9123 & \textbf{0.2132} & 0.7784 & 1.6028 & \textbf{0.0279} & \underline{0.2784} & 0.2013 & 0.0000 & 0.0586 \\        
        WS & 0.8293 & 0.9109 & \textbf{0.1064} & 1.3336 & 0.1307 & 1.1000 & - & - & - \\
        MMSB & 1.6263 & 1.3005 & 0.2668 & - & - & - & 0.5927 & 0.0000 & 0.0217 \\
        Kronecker & 1.0575 & 1.3005 & 0.2460 & 1.7002 & 0.1310 & 1.2000 & 1.0424 & 0.0000 & 0.0246 \\
        \hline
        Ours & \textbf{0.2728} & \underline{0.5245} & \underline{0.1635} & \textbf{0.3229} & 1.3575 & \textbf{0.0231} & \textbf{0.0013} & \textbf{0.0000} & \textbf{0.0003} \\
        \hline
    \end{tabular}
    \label{tab:main}
\end{table*}
Table~\ref{tab:main} shows the comparison results of our method to baselines on all 12 datasets.
The best and second scores are in \textbf{bold} and \underline{underlined}, respectively.
Absences in the table, marked by -, denote the failure of generation (e.g., self-loop, no edges, etc.) or out of time (OOT).
Generally, our method achieves better generation results than traditional works.
Although some scores of particular datasets are not the best, we are still in the top two performances, and taking all metrics together, we can provide significant improvement.

In social networks, our method fails in \textit{clus.} of REDDIT-BINARY.
We believe that the significant discrepancy between the high clustering characteristics and scale-free assumption causes the failure.
The same degradation occurs in synthetic datasets, GRID, where the regular lattice structures contradict scale-free principles.
Finally, in bioinformatics, our method shows a generally weaker performance compared to other dataset categories.
We think that the prevalence of loops (e.g. benzene rings) effects more to the generation performance by observing the superior generation results of WS which is on the basis of ring structures.

Though limited resources setting restricts the ability of learning models, we also provide extra experiments, comparing our method with deep generative models and large language models.
Please refer to Appendix E.2 and E.3 to see more details.
\subsection{Parameter Studies}
\begin{figure*}[!t]
    \centering
    \includegraphics[width=0.95\textwidth]{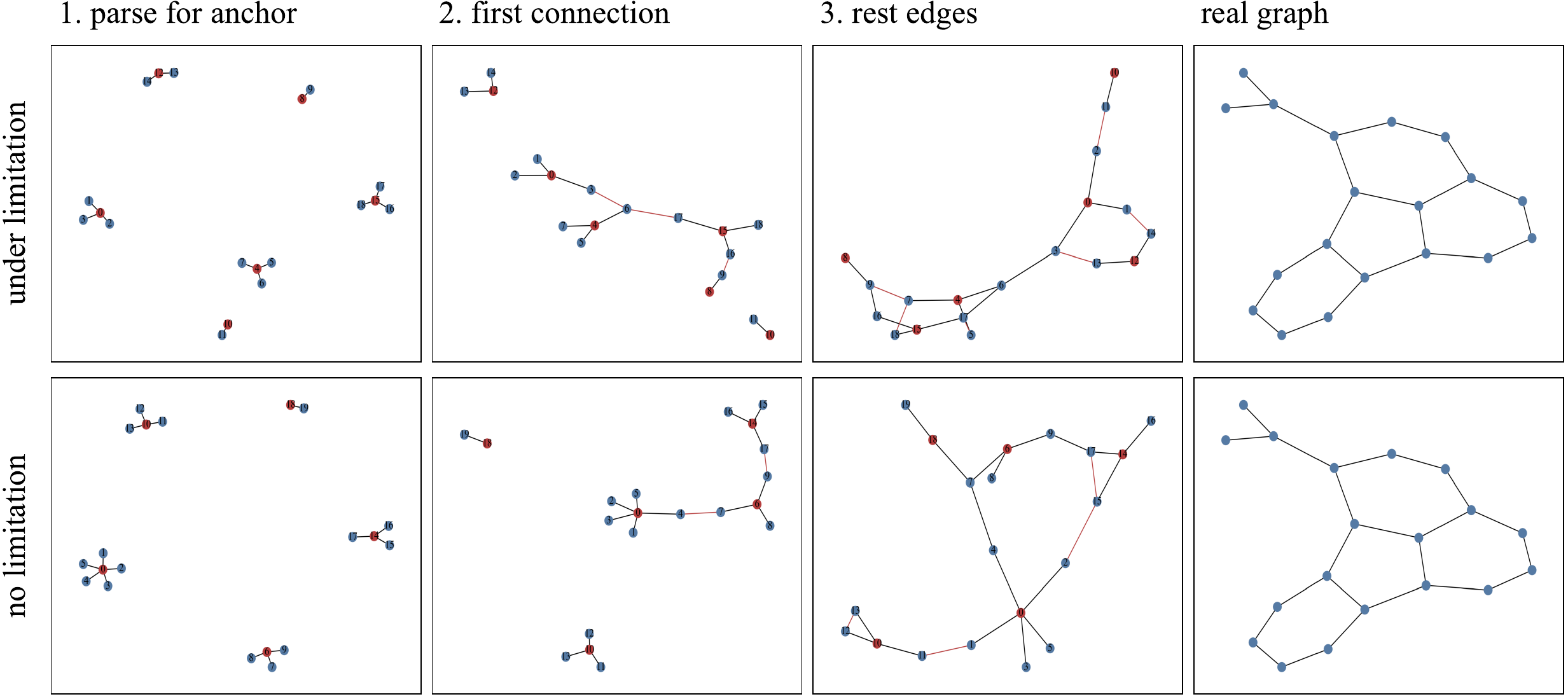} 
    \caption{Generation steps with/without $d_{max}$ limitation. Anchor nodes and edges to add in each step are in red.}
    \label{fig:maxlim}
\end{figure*}

There are two parameters in our method: one is $d_{max}$, and the other is $k$.
They both determine how tolerant the generation process is of particular graph cases.
The former limits the maximum degree of each node and mainly affects the selection of anchor nodes.
Figure~\ref{fig:maxlim} illustrates the comparison of generation steps with and without $d_{max}$ limitation.
For example, without $d_{max} = 3$, our method may generate $\mathcal{G}_{sub}=(6, 5)$ in the first stage, which finally impacts the following edge selection.
The star graph formed by the anchor node exceeding the $d_{max}$ degree constraint will change the properties of the generated graph, creating structures out of distribution.

The latter is essentially a truncation.
Since a connected graph has no node with 0 degrees, we can assume that nodes with a degree probability lower than $P(0|\bar{D})$ do not exist.
We define $k$ to be the first positive integer satisfying $P(k|\bar{D}) < P(0|\bar{D})$, which controls whether small probability nodes occur.
Table~\ref{tab:maxlim} shows the ablation experiments for the two parameters above, reporting the average MMD scores of the four datasets under each category.
Results show that both parameters are positive to the performance, and $d_{max}$ shows a more competitive contribution than $k$.
 \begin{table*}[!t]
    \centering
    \caption{Ablation of two parameters on all three dataset categories.}
    \begin{tabular}{cc|ccc|ccc|ccc}
        \hline
        \multirow{2}{*}{} & & \multicolumn{3}{c}{Bioinformatics \& Molecules} & \multicolumn{3}{c}{Social Networks} & \multicolumn{3}{c}{Synthetic} \\
        $d_{max}$ & $k$ & deg. & clus. & orbit & deg. & clus. & orbit & deg. & clus. & orbit \\
        \hline
        $\sqrt{}$ & $\sqrt{}$ & \textbf{0.1309} & \textbf{0.2673} & \textbf{0.0731} & \textbf{0.0308} & \textbf{0.3904} & \textbf{0.0626} & \textbf{0.0924} & \textbf{0.1498} & \textbf{0.0185} \\
        $\sqrt{}$ & $\times$ & \underline{0.1368} & \underline{0.3617} & \underline{0.1269} & \underline{0.0362} & \underline{0.4870} & \underline{0.0837} & \underline{0.1057} & \underline{0.1804} & \underline{0.0323} \\
        $\times$ & $\sqrt{}$ & 0.1971 & 0.5051 & 0.1397 & 0.1901 & 0.5739 & 0.2157 & 0.1874 & 0.1954 & 0.1482 \\
        $\times$ & $\times$ & 0.2173 & 0.5483 & 0.1438 & 0.1992 & 0.5959 & 0.2380 & 0.1926 & 0.1954 & 0.1482 \\
        \hline
    \end{tabular}
    \label{tab:maxlim}
\end{table*}
\subsection{Graph Generation under Various Distributions}
\begin{table*}[!t]
    \centering
    \caption{MMD evaluation under various distributions.}
    \begin{tabular}{c|ccc|ccc|ccc}
        \hline
        \multirow{2}{*}{} & \multicolumn{3}{c}{ENZYMES} & \multicolumn{3}{c}{deezer\_ego\_nets} & \multicolumn{3}{c}{CLUS} \\
        Methods & deg. & clus. & orbit & deg. & clus. & orbit & deg. & clus. & orbit \\
        \hline
        Uniform & 0.5699 & 1.1729 & 0.8841 & 0.1395 & 0.2781 & 0.0696 & 0.1656 & \underline{0.1946} & 0.0920 \\
        Normal & 1.3402 & 1.0053 & 1.0025 & 0.3432 & 0.4294 & 0.1433 & 0.7931 & 0.6799 & 0.4577 \\
        Exponential & 0.8958 & 1.0560 & 1.1191 & 0.1274 & 0.1794 & 0.0787 & 0.1614 & 0.1975 & 0.1666 \\
        Gamma & 0.9899 & \underline{0.8145} & \underline{0.4089} & \underline{0.0460} & \underline{0.0915} & \underline{0.0097} & 0.2800 & 0.4155 & 0.1168 \\
        Pareto & \underline{0.1898} & 0.8487 & 0.4196 & 0.1417 & 0.3587 & 0.0757 & \underline{0.1480} & 0.3558 & \underline{0.0758} \\
        \hline
        Ours & \textbf{0.1811} & \textbf{0.4621} & \textbf{0.1271} & \textbf{0.0371} & \textbf{0.0668} & \textbf{0.0026} & \textbf{0.0840} & \textbf{0.1911} & \textbf{0.0509} \\
        \hline
        \hline
        \multirow{2}{*}{} & \multicolumn{3}{c}{MUTAG} & \multicolumn{3}{c}{IMDB-BINARY} & \multicolumn{3}{c}{EGO} \\
        Methods & deg. & clus. & orbit & deg. & clus. & orbit & deg. & clus. & orbit \\
        \hline
        Uniform & 0.3167 & 0.3888 & 0.1086 & 0.1479 & 0.7773 & 0.2317 & 0.9074 & 0.0000 & 0.5520 \\
        Normal & 0.5351 & 0.5004 & 0.1450 & 0.5009 & 0.8092 & 0.2353 & 1.1860 & 0.0000 & 0.8863 \\
        Exponential & 0.6231 & 0.2110 & 0.1739 & 0.2421 & 0.7752 & 0.2321 & 1.1360 & 0.0000 & 0.6298 \\
        Gamma & \underline{0.1975} & \underline{0.1198} & \underline{0.0164} & 0.3261 & 0.9085 & \underline{0.1509} & \underline{0.6218} & 0.0000 & \underline{0.0701} \\
        Pareto & 0.6135 & 0.6105 & 0.2943 & \underline{0.1214} & \underline{0.7296} & 0.2264 & 1.1181 & 0.0000 & 0.9514 \\
        \hline
        Ours & \textbf{0.0355} & \textbf{0.0561} & \textbf{0.0014} & \textbf{0.0393} & \textbf{0.7161} & \textbf{0.1436} & \textbf{0.0382} & \textbf{0.0000} & \textbf{0.0080} \\
        \hline
        \hline
        \multirow{2}{*}{} & \multicolumn{3}{c}{NCI1} & \multicolumn{3}{c}{IMDB-MULTI} & \multicolumn{3}{c}{GRID} \\
        Methods & deg. & clus. & orbit & deg. & clus. & orbit & deg. & clus. & orbit \\
        \hline
        Uniform & 0.5569 & 0.1396 & 0.3430 & 0.0892 & 0.4853 & 0.1013 & 0.5060 & 1.4178 & \underline{0.1250} \\
        Normal & 0.7984 & 0.2309 & 0.7473 & 0.2553 & \underline{0.4401} & \underline{0.0920} & 1.2460 & 1.1215 & 0.5624 \\
        Exponential & 0.6656 & 0.1147 & 0.4386 & 0.1404 & 0.5420 & 0.1047 & 1.0376 & 1.3562 & 0.4547 \\
        Gamma & \underline{0.3090} & \underline{0.0359} & \underline{0.0250} & 0.2247 & 0.4844 & 0.1081 & 0.6368 & \underline{1.0250} & 0.2682 \\
        Pareto & 0.8346 & 0.2959 & 0.7700 & \underline{0.0509} & 0.5417 & 0.1036 & \underline{0.3485} & 1.1160 & 0.1606 \\
        \hline
        Ours & \textbf{0.0340} & \textbf{0.0263} & \textbf{0.0002} & \textbf{0.0160} & \textbf{0.3882} & \textbf{0.0416} & \textbf{0.2459} & \textbf{0.4081} & \textbf{0.0148} \\
        \hline
        \hline
        \multirow{2}{*}{} & \multicolumn{3}{c}{PROTEINS} & \multicolumn{3}{c}{REDDIT-BINARY} & \multicolumn{3}{c}{TREE} \\
        Methods & deg. & clus. & orbit & deg. & clus. & orbit & deg. & clus. & orbit \\
        \hline
        Uniform & 0.4829 & 0.6092 & 0.3512 & \underline{0.3513} & \underline{0.2779} & 0.1046 & 0.1042 & 0.0000 & 0.0382 \\
        Normal & 1.1046 & 0.6546 & 0.5062 & 0.5060 & 1.4178 & 0.1250 & 0.3458 & 0.0000 & 0.1303 \\
        Exponential & 0.6797 & 0.5711 & 0.4015 & 0.5148 & 0.4749 & 0.2000 & 0.1684 & 0.0000 & 0.0529 \\
        Gamma & 0.6342 & 0.5768 & 0.1909 & 0.6295 & \textbf{0.1239} & \underline{0.0240} & 0.0995 & 0.0000 & \underline{0.0019} \\
        Pareto & \underline{0.2734} & \underline{0.5335} & \underline{0.1853} & 0.4491 & 0.5945 & 0.1038 & \underline{0.0276} & 0.0000 & 0.0305 \\
        \hline
        Ours & \textbf{0.2728} & \textbf{0.5245} & \textbf{0.1635} & \textbf{0.3229} & 1.3575 & \textbf{0.0231} & \textbf{0.0013} & \textbf{0.0000} & \textbf{0.0003} \\
        \hline
    \end{tabular}
    \label{tab:distribution}
\end{table*}
We conduct a distribution ablation experiment to demonstrate the superiority of our method, replacing the Poisson distribution with other distributions.
We pick five distributions: Uniform, Gaussian (standard normal), Exponential, Gamma, and Pareto.
The detailed parameter settings are in Appendix E.1.
For a possible degree $x=d$, we consider the cumulative distribution function (CDF) as an approximation of its probability, where $P\{x=d\}=P\{d-1<x\leq d\}$.
Ablation results are in Table~\ref{tab:distribution}.
Through experiments, we can draw two enlightened conclusions.
First, our method achieves the best results, indicating that the observation of the scale-free property of which this paper is the basis is valid.
Next, we observe from the underlined results that different prior distributions fit datasets in disparate categories.
Generally, Uniform and Normal are more suitable for graphs with high clustering, making edge selection of nodes in clusters tend to be consistent.
Other distributions with approximate scale-free properties can also achieve competitive results under our hierarchical design, but are overall weaker than our method.
In conclusion, although different datasets and distributions have their specific features, our method is relatively one of the optimal choices in the scenario of unknown distribution.
\subsection{Comparison with Real Degree Distributions}
\begin{figure*}[!t]
    \centering
    \begin{subfigure}
        \centering
        \includegraphics[width=0.8\textwidth]{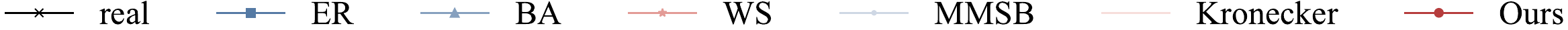} 
    \end{subfigure}
    \begin{subfigure}
        \centering
        \includegraphics[width=0.23\textwidth]{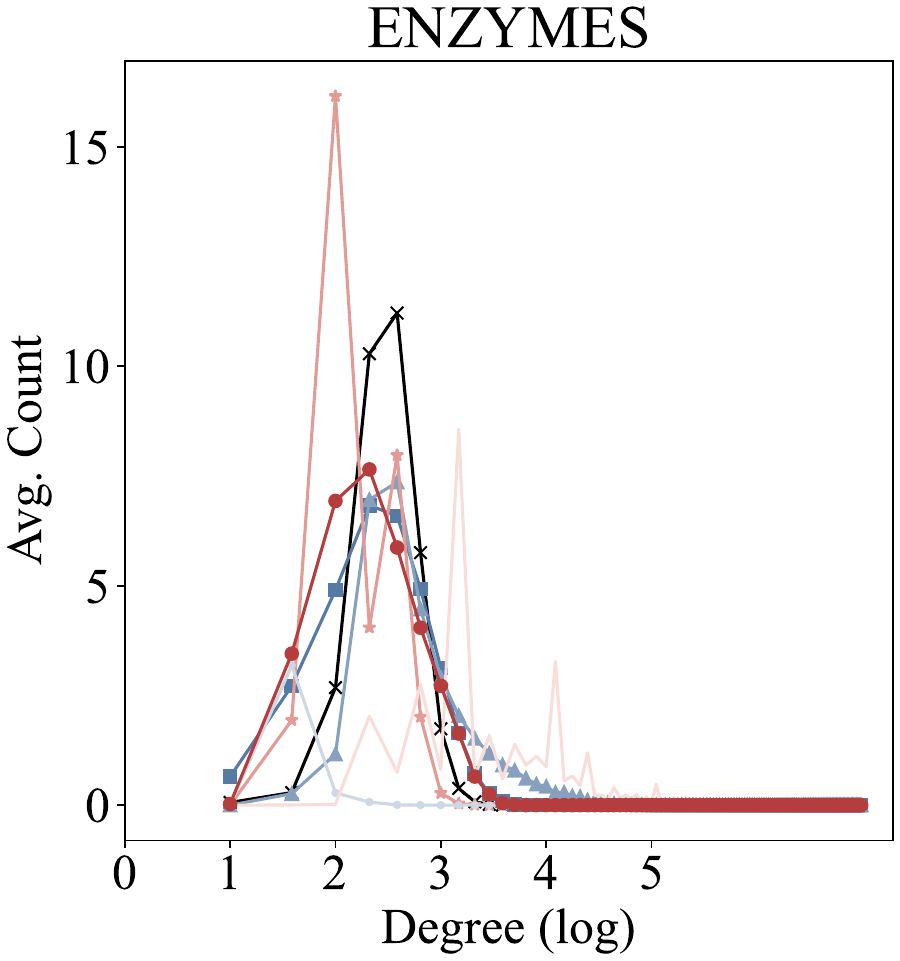}
    \end{subfigure}
    \begin{subfigure}
        \centering
        \includegraphics[width=0.238\textwidth]{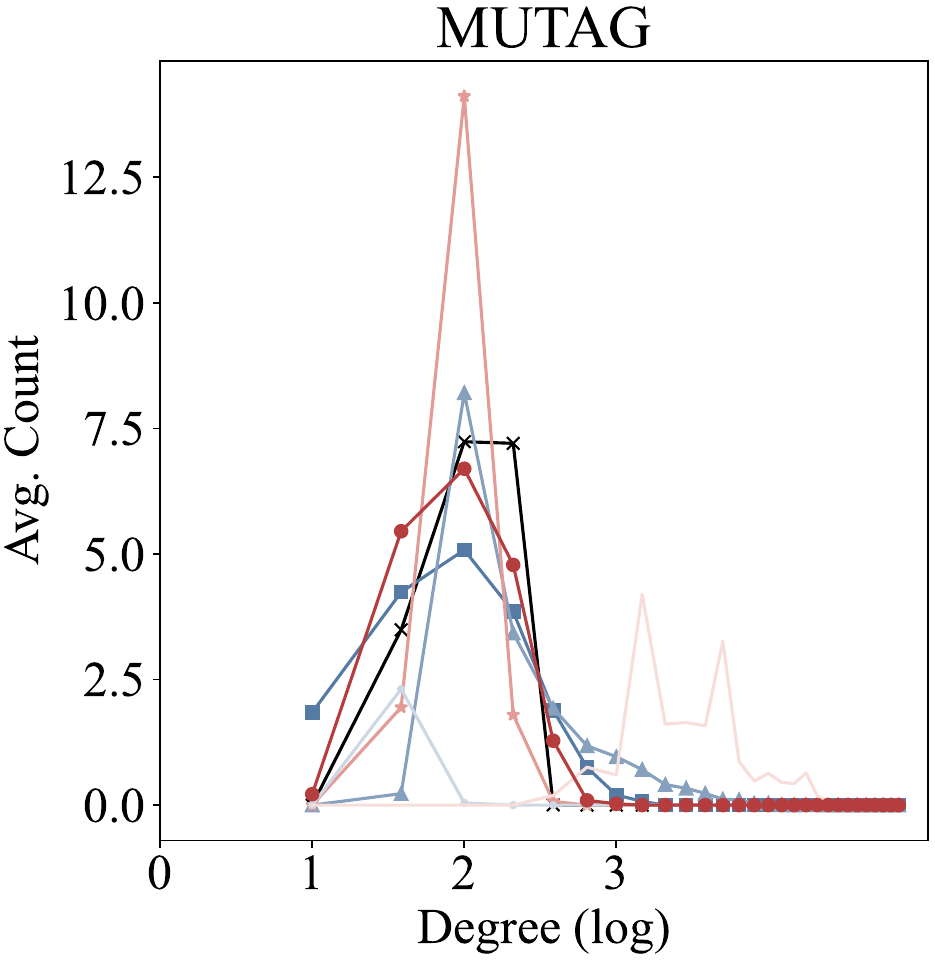}
    \end{subfigure}
    \begin{subfigure}
        \centering
        \includegraphics[width=0.23\textwidth]{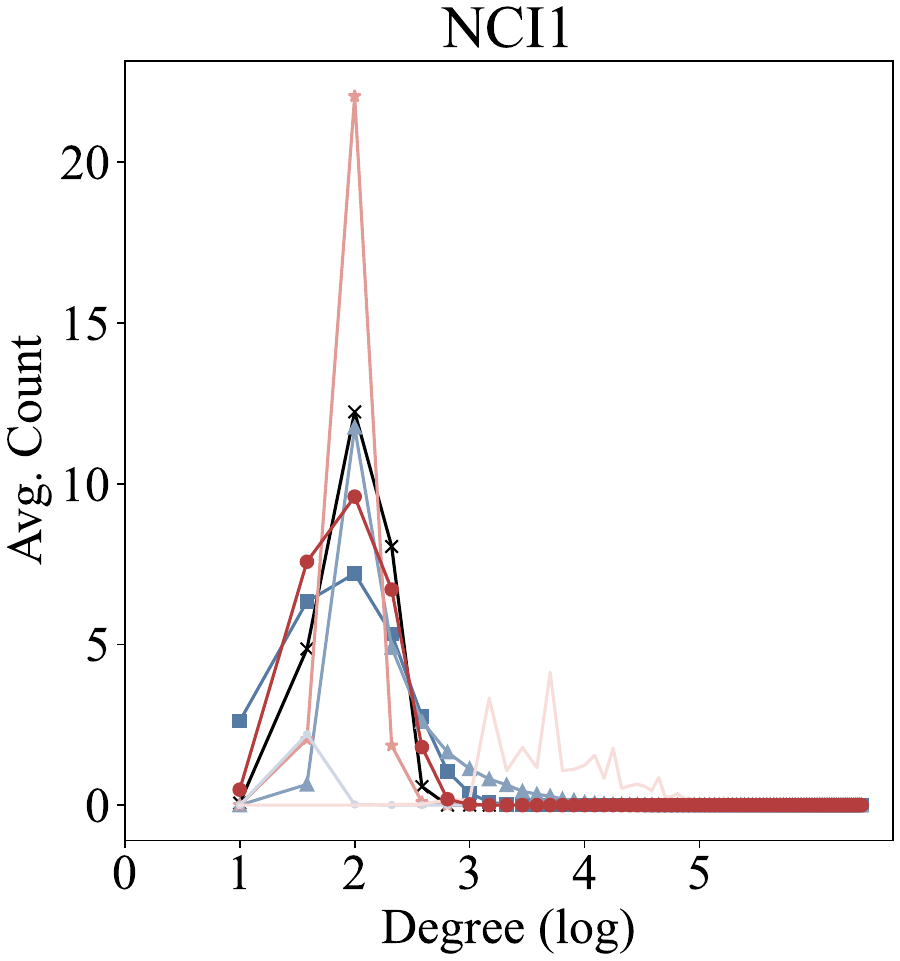}
    \end{subfigure}
    \begin{subfigure}
        \centering
        \includegraphics[width=0.23\textwidth]{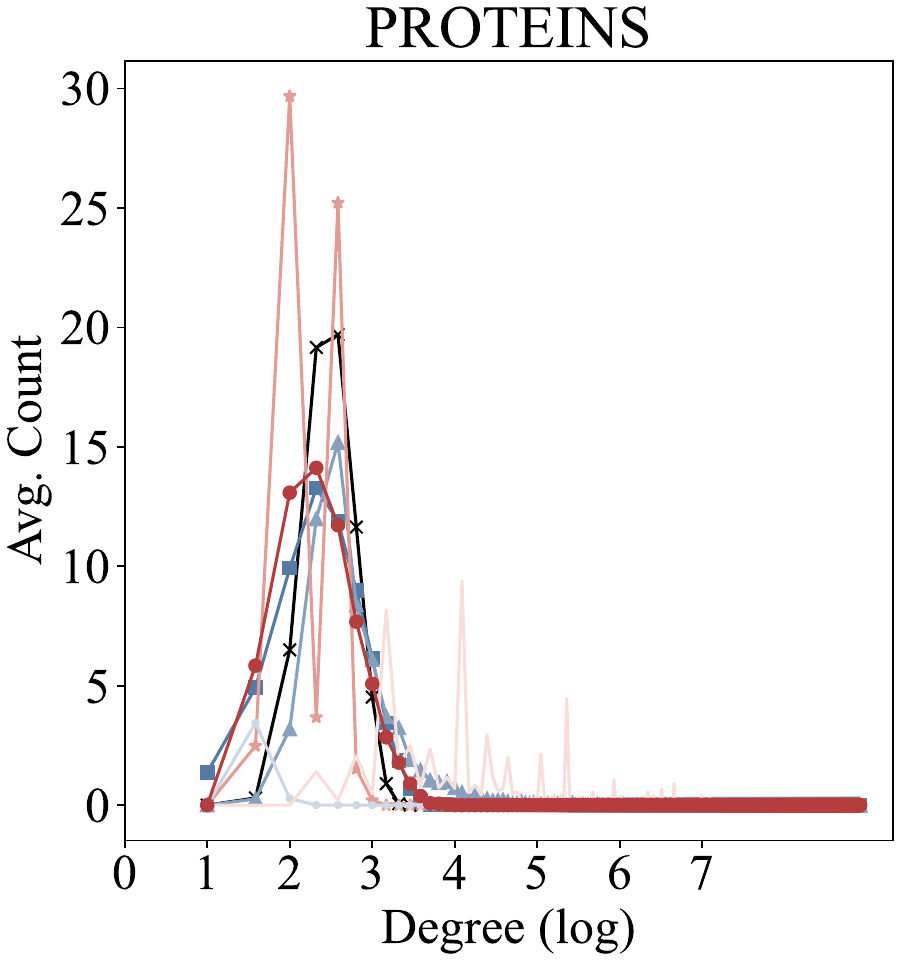}
    \end{subfigure}
    \vspace{0.01\textwidth}
    \begin{subfigure}
        \centering
        \includegraphics[width=0.232\textwidth]{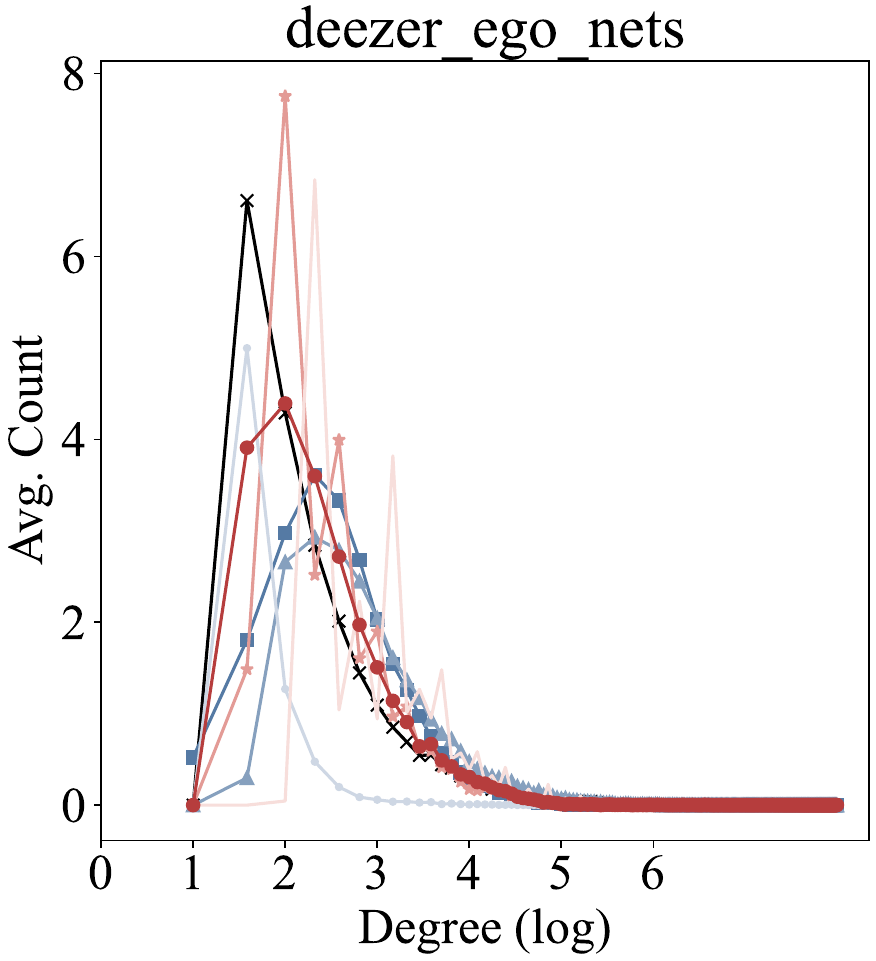}
    \end{subfigure}
    \begin{subfigure}
        \centering
        \includegraphics[width=0.238\textwidth]{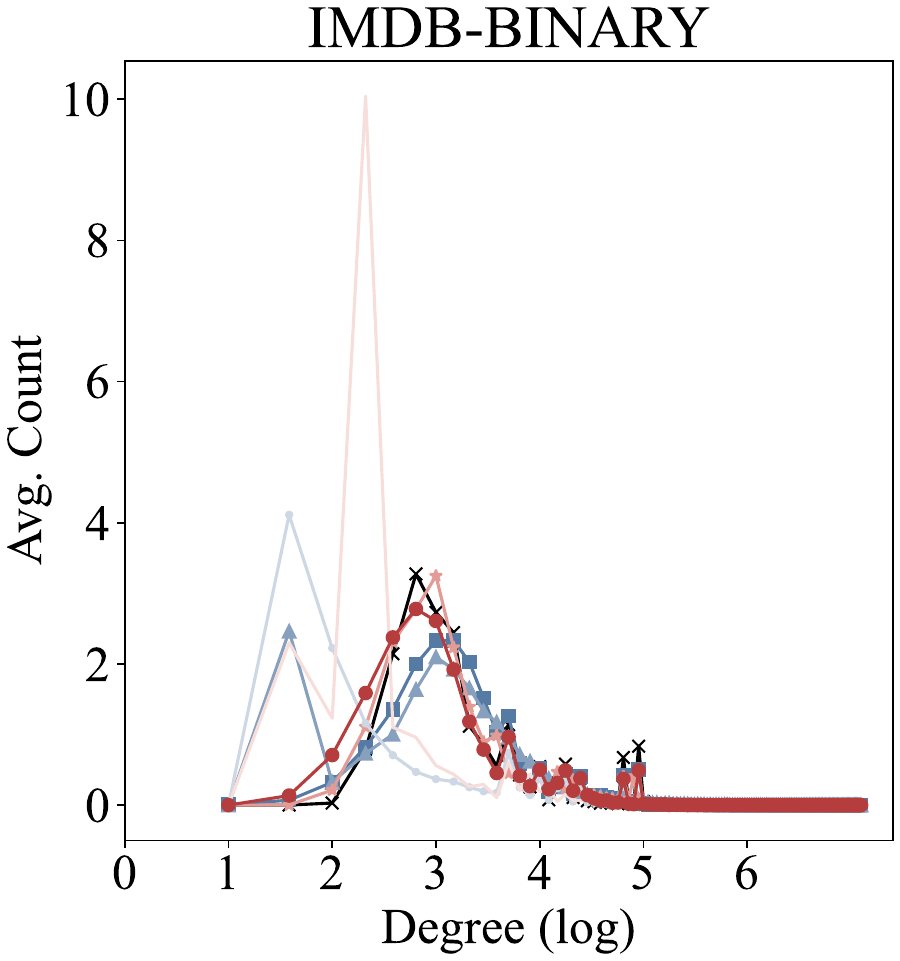}
    \end{subfigure}
    \begin{subfigure}
        \centering
        \includegraphics[width=0.23\textwidth]{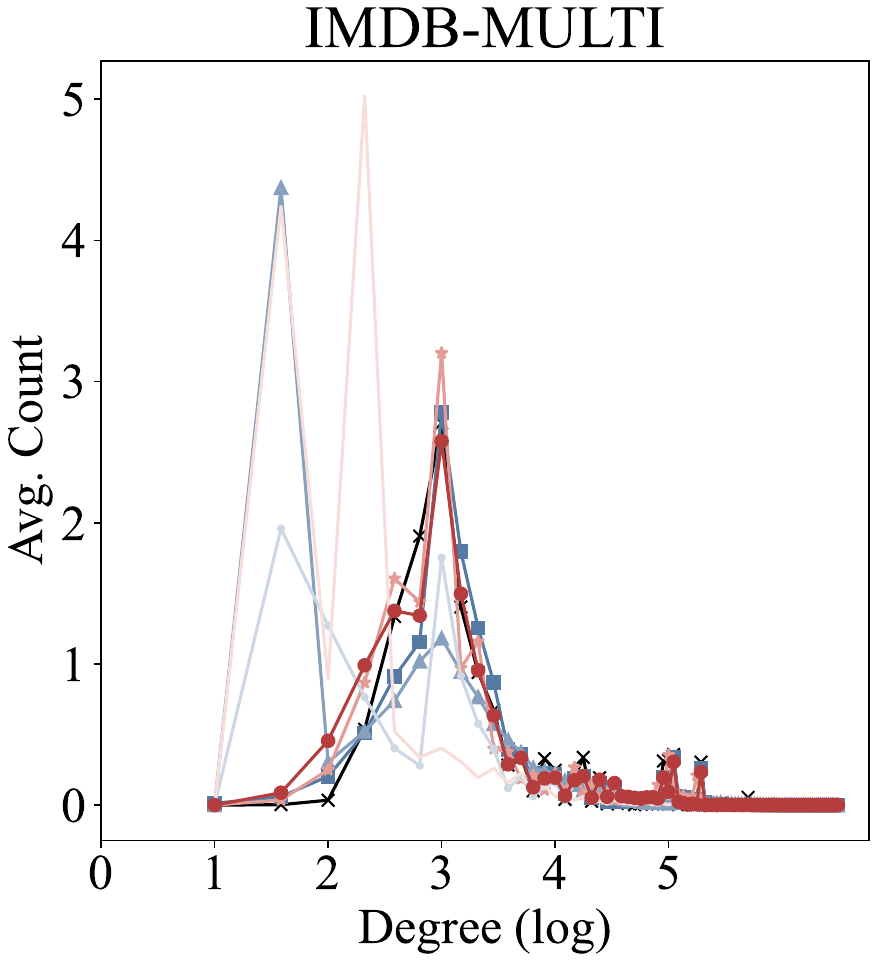}
    \end{subfigure}
    \begin{subfigure}
        \centering
        \includegraphics[width=0.24\textwidth]{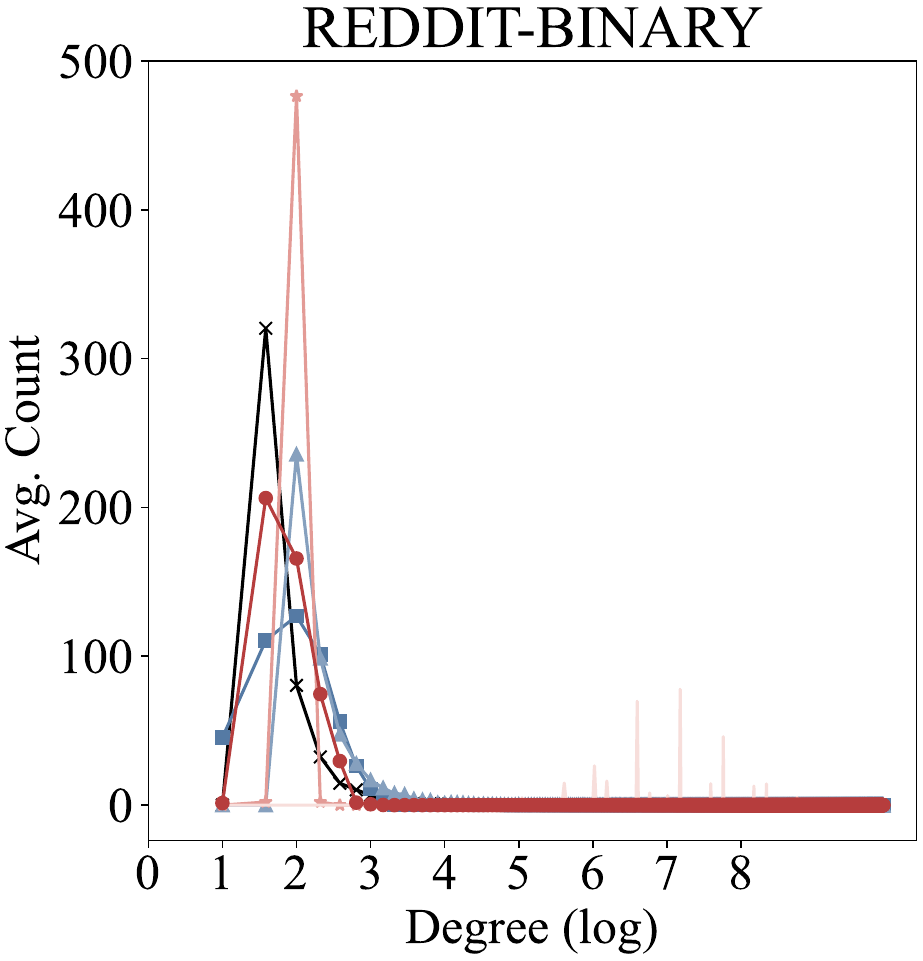}
    \end{subfigure}
    \vspace{0.01\textwidth}
    \begin{subfigure}
        \centering
        \includegraphics[width=0.23\textwidth]{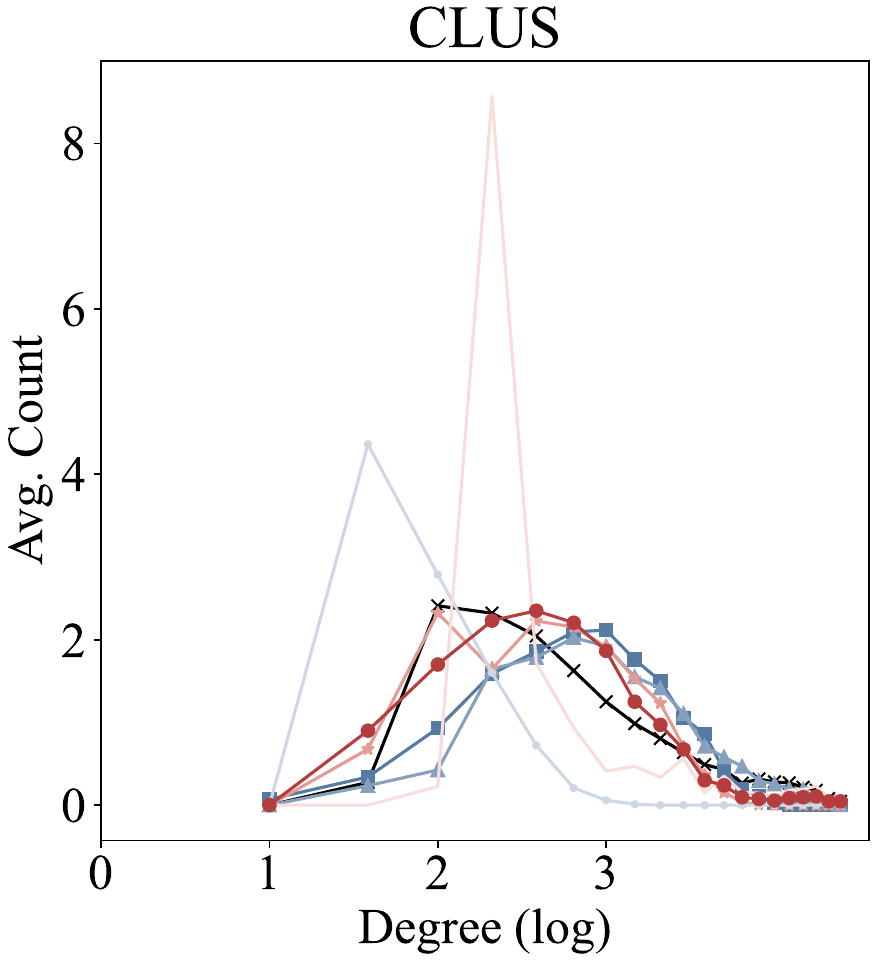}
    \end{subfigure}
    \begin{subfigure}
        \centering
        \includegraphics[width=0.24\textwidth]{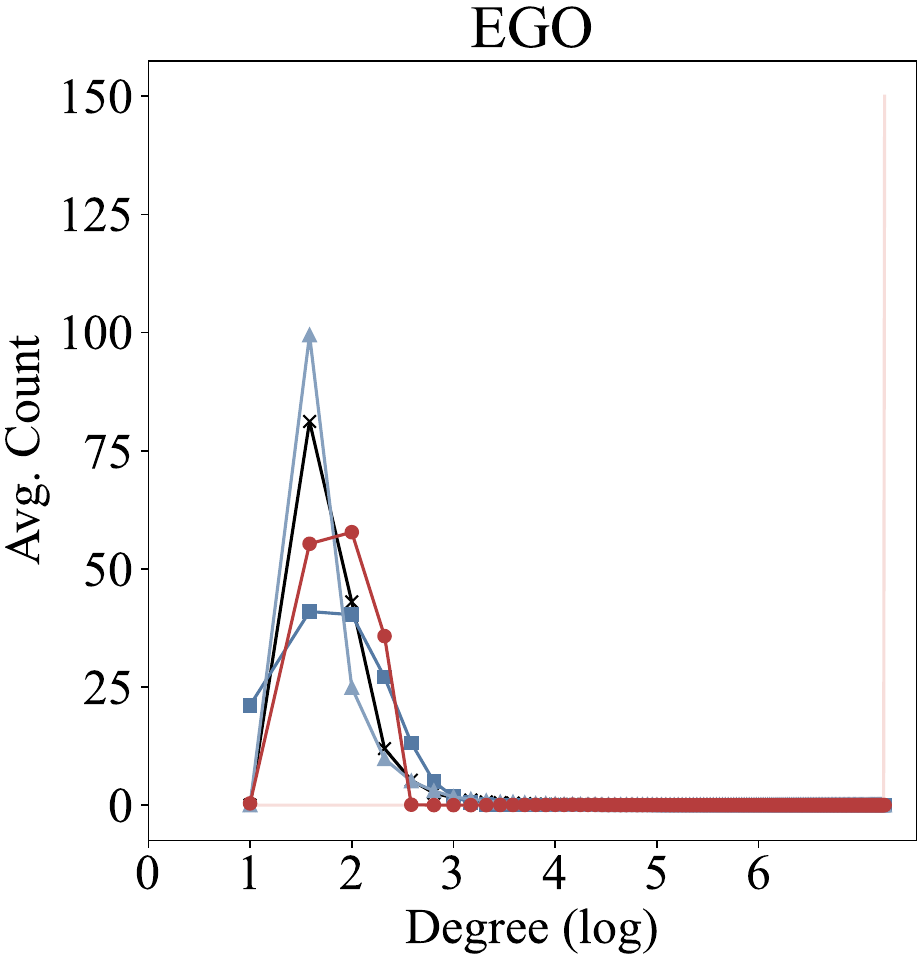}
    \end{subfigure}
    \begin{subfigure}
        \centering
        \includegraphics[width=0.232\textwidth]{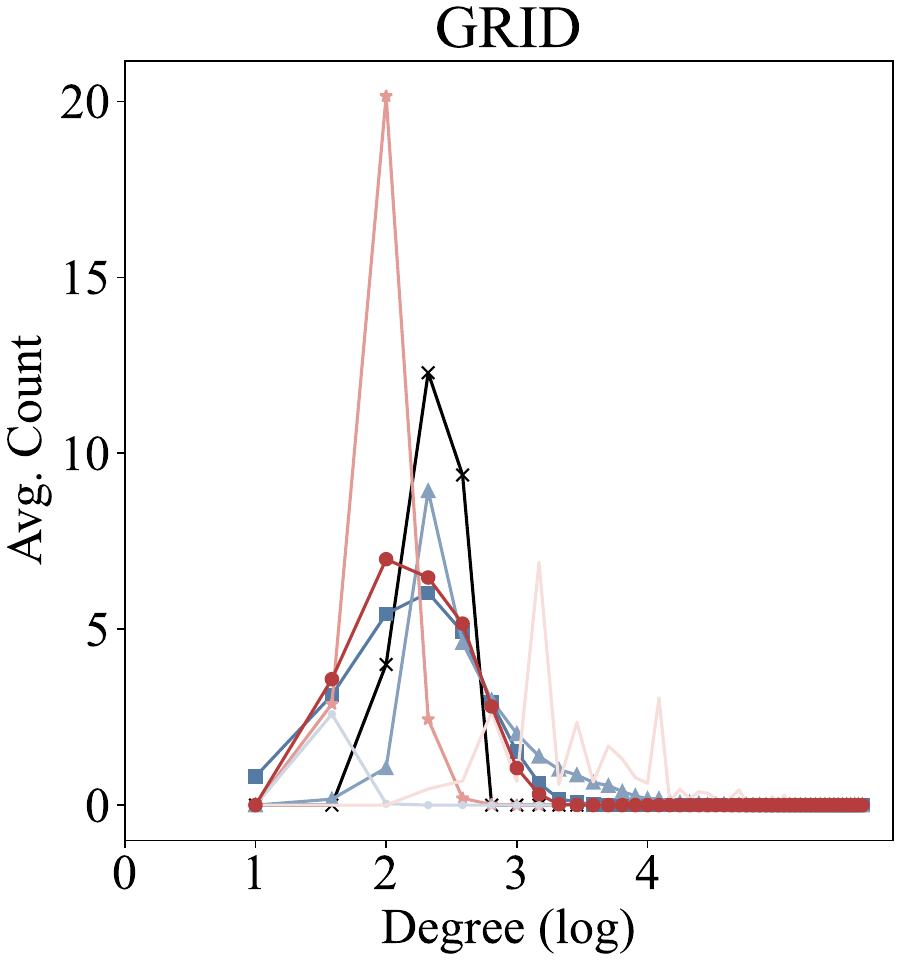}
    \end{subfigure}
    \begin{subfigure}
        \centering
        \includegraphics[width=0.228\textwidth]{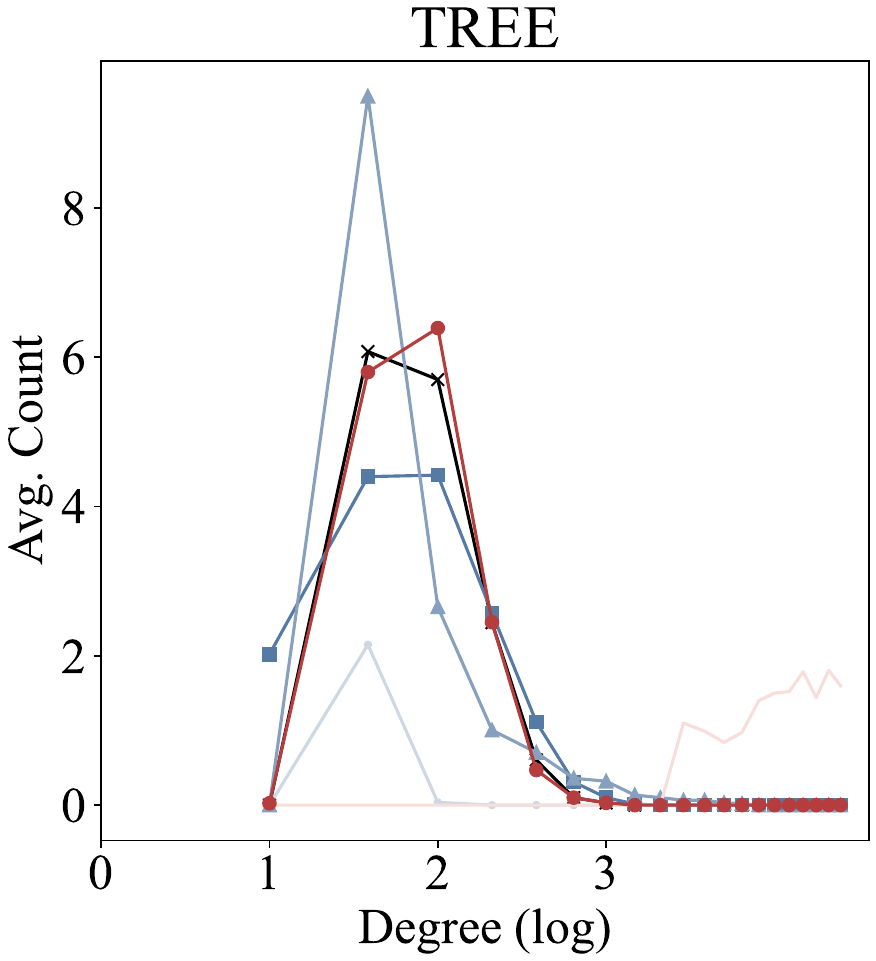}
    \end{subfigure}
    \caption{Degree distribution comparisons between generated and real-world graphs.}
    \label{fig:degree}
\end{figure*}
\begin{figure*}[!ht]
    \centering
    \includegraphics[width=\linewidth]{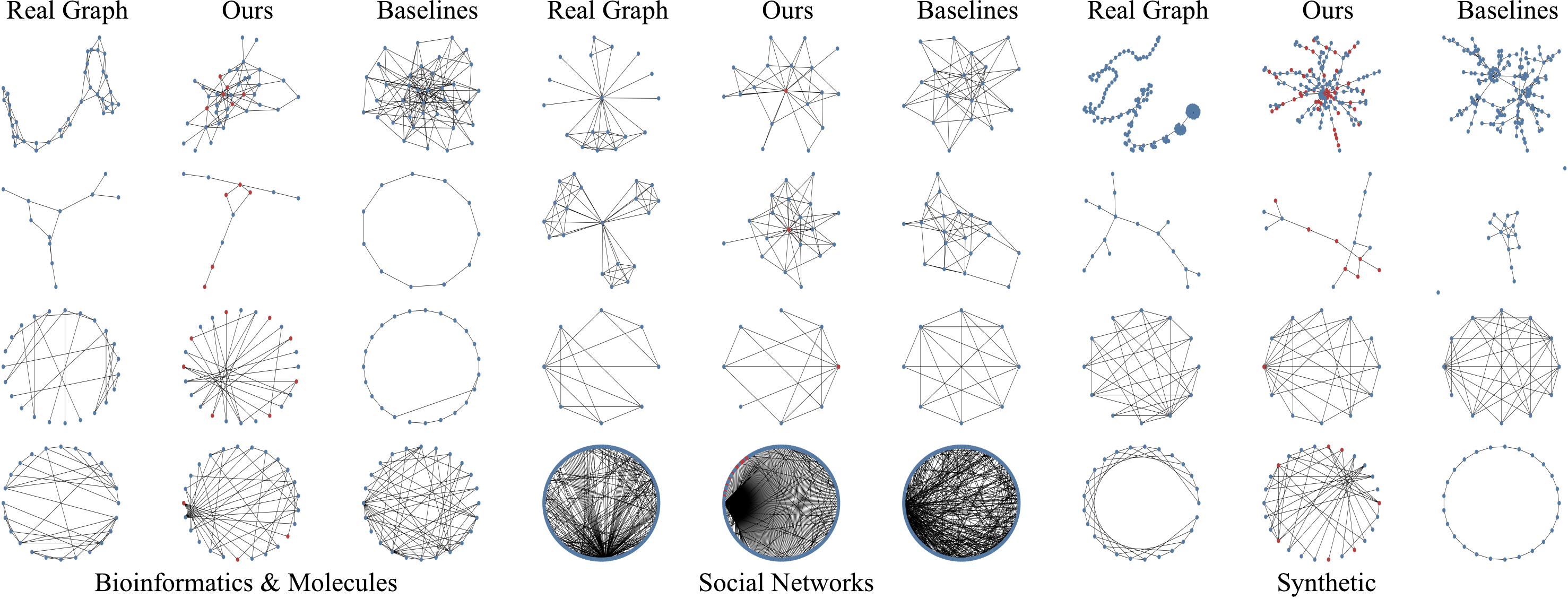}
    \caption{Visualization of graphs of all categories. They are (from left to right, each consisting of three columns) bioinformatics \& molecules, social networks, and synthetic graphs. Within each category, the first column is from the dataset, and the third is from baselines achieving competitive results. The second column is from our method, where anchor nodes are in red. We use various layouts for each row to represent different datasets.}
    \label{fig:visual}
\end{figure*}
Until now, we have restricted access to the ground truth distribution.
We are so curious about the difference between the degree distribution of the generated and real-world graphs in such a setting that we design experiments to collect those statistics.
Figure~\ref{fig:degree} shows the results from baselines and our method.
The x-axis represents the node degree with a truncation to count only the nodes with degrees less than 20, while the y-axis represents the average over all graphs in the datasets.
Experimental results show that our method can generate graphs close to the ground truth distribution.
Despite the differences between categories, we can see the scale-free characteristics of all datasets in Figure~\ref{fig:degree}.
This property prevents the occurrence of extra degrees when generating graphs.
Another observation is that for cases with more complicated distributions, such as four social networks in the middle row, our method can also fit the fluctuation of the plots.
For example, on IMDB-BINARY and IMDB-MULTI, our plots (red solid circle) almost coincide with the plots of ground truth (black cross).
Some baselines may have a better coincidence in a particular degree interval than our method but severely fluctuate when dealing with all datasets along all degrees.
\subsection{Graph Visualization}
We visualize the graphs generated by baselines and our method.
Figure~\ref{fig:visual} illustrates the results.
Our method can maximally recover the unique structures that may appear under the real-world distribution, such as loops (row 2, column 2), cluster centers (row 1, column 5), chains (row 2, column 8), etc.
Since the ground truth distribution is unknown, there are still limitations in fully generating structures rich in semantics, for example, benzene rings (loops containing six nodes) in bioinformatics, standard 2D grids, etc.
Our method best succeeds data with similar scale-free properties, where closer proximity means more prior knowledge of the distribution.
Meanwhile, we reduce the generation of invalid structures, such as plain graphs (row 2, column 3), etc.

\section{Conclusion}
In this paper, we design a hierarchical scale-free graph generator under limited resources.
We consider the scale-free property of graphs as an invariant feature and propose a two-stage generation strategy.
We split the original set of nodes into multiple substructures by sampling anchor nodes, ensuring scale-free distribution and local clustering property.
We also design a degree mixing distribution to sample the remaining edges, cooperating with selected thresholds, which can adjust the algorithm tolerance to exotic structures.  
The experimental results on 12 datasets in three categories show that our method can fit the ground truth distribution and generate higher-quality graphs than SOTAs.


\begin{thebibliography}{88}
\bibitem{ref:social2021} C. Wang, B. Wang, B. Huang, S. Song, and Z. Li, ``FastSGG: Efficient Social Graph Generation Using a Degree Distribution Generation Model,'' in \textit{ICDE}, 2021.
\bibitem{ref:social2022} S. Xiang, D. Cheng, J. Zhang, Z. Ma, X. Wang, and Y. Zhang, ``Efficient Learning-based Community-Preserving Graph Generation,'' in \textit{ICDE}, 2022.
\bibitem{ref:social2024} L. Zhang, D. Zhou, H. Tong, J. Xu, Y. Zhu, and J. He, ``FairGen: Towards Fair Graph Generation,'' in \textit{ICDE}, 2024.
\bibitem{ref:bio2019} H. Huang, L. Sun, B. Du, and W. Lv, ``Conditional Diffusion Based on Discrete Graph Structures for Molecular Graph Generation,'' in \textit{AAAI}, 2023.
\bibitem{ref:bio2023} Y. Zhu, Z. Ouyang, B. Liao, J. Wu, Y. Wu, C. Hsieh, T. Hou, and J. Wu, ``MolHF: A Hierarchical Normalizing Flow for Molecular Graph Generation,'' in \textit{IJCAI}, 2023.
\bibitem{ref:bio2024} G. Wei, Y. Huang, C. Duan, Y. Song, and Y. Du, ``Navigating Chemical Space with Latent Flows,'' in \textit{NeurIPS}, 2024.
\bibitem{ref:ss2022} H. Huang, L. Sun, B. Du, Y. Fu, and W. Lv, ``GraphGDP: Generative Diffusion Processes for Permutation Invariant Graph Generation,'' in \textit{ICDM}, 2022.
\bibitem{ref:ss2023} N. Gruver, S. D. Stanton, N. C. Frey, T. G. J. Rudner, I. Hotzel, J. L. Vanasse, A. Rajpal, K. Cho, and A. G. Wilson, ``Protein Design with Guided Discrete Diffusion,'' in \textit{NeurIPS}, 2023.
\bibitem{ref:ss2024} Y. Du, M. Plainer, R. Brekelmans, C. Duan, F. No\'e, C. P. Gomes, A. A. Guzik, and K. Neklyudov, ``Doob's Lagrangian: A Sample-Efficient Variational Approach to Transition Path Sampling,'' in \textit{NeurIPS}, 2024.
\bibitem{ref:DGM} X. Guo, Y. Du, and L. Zhao, ``Deep Generative Models for Spatial Networks,'' in \textit{KDD}, 2021.
\bibitem{ref:trans2019} C. Zhang, X. Lyu, and Z. Tang, ``TGG: Transferable Graph Generation for Zero-shot and Few-shot Learning,'' in \textit{MM}, 2019.
\bibitem{ref:trans2023} S. Zhang, T. Li, S. Hui, G. Li, Y. Liang, L. Yu, D. Jin, and Y. Li, ``Deep Transfer Learning for City-scale Cellular Traffic Generation through Urban Knowledge Graph,'' in \textit{KDD}, 2023.
\bibitem{ref:trans2024} J. Yang, X. Jiang, Y. Guo, L. T. Yang, and J. Yang, ``Generalize to Fully Unseen Graphs: Learn Transferable Hyper-Relation Structures for Inductive Link Prediction,'' in \textit{MM}, 2024.
\bibitem{ref:DS} Bernardino Romera-Paredes and Philip Torr, ``An embarrassingly simple approach to zero-shot learning,'' in \textit{ICML}, 2015.
\bibitem{ref:LinkBench} T. G. Armstrong, V. Ponnekanti, D. Borthakur, and M. Callaghan, ``LinkBench: A database benchmark based on the Facebook social graph,'' in \textit{SIGMOD}, 2013.
\bibitem{ref:LinkGen} A. K. Joshi, P. Hitzler, and G. Dong, ``LinkGen: Multipurpose linked data generator,'' in \textit{ISWC}, 2016.
\bibitem{ref:gMark} G. Bagan, A. Bonifati, R. Ciucanu, G. Fletcher, A. Lemay, and N. Advokaat, ``gMark: Schema-driven generation of graphs and queries,'' in \textit{TKDE}, 2016.
\bibitem{ref:adj2020} B. Samanta, A. De, G. Jana, V. Gomez, P. K. Chattaraj, N. Ganguly, and M. G. Rodriguez, ``NeVAE: A Deep Generative Model for Molecular Graphs,'' in \textit{JMLR}, 2020.
\bibitem{ref:adj2021} Y. Du, Y. Wang, F. Alam, Y. Lu, X. Guo, L. Zhao, and A. Shehu, ``Deep Latent-Variable Models for Controllable Molecule Generation,'' in \textit{BIBM}, 2021.
\bibitem{ref:adj2022} Y. Du, X. Guo, A. Shehu, and L. Zhao, ``Interpretable Molecular Graph Generation via Monotonic Constraints,'' in \textit{SDM}, 2022.
\bibitem{ref:adj2023} Y. Liu, X. Ao, F. Feng, Y. Ma, K. Li, T. Chua, and Q. He, ``FLOOD: A Flexible Invariant Learning Framework for Out-of-Distribution Generalization on Graphs,'' in \textit{KDD}, 2023.
\bibitem{ref:seq2022} Y. Liu, L. Zou, and Z. Wei, ``Building Graphs at Scale via Sequence of Edges: Model and Generation Algorithms (Extended Abstract),'' in \textit{ICDE}, 2022.
\bibitem{ref:seq2024} S. Zheng, C. Wang, C. Wu, Y. Lou, H. Feng, and X. Yang, ``Temporal Graph Generation Featuring Time-Bound Communities,'' in \textit{ICDE}, 2024.
\bibitem{ref:seman2022} Clement Vignac and Pascal Frossard, ``Top-N: Equivariant Set and Graph Generation without Exchangeability,'' in \textit{ICLR}, 2022.
\bibitem{ref:seman2023} B. Qiang, Y. Song, M. Xu, J. Gong, B. Gao, H. Zhou, W. Ma, and Y. Lan, ``Coarse-to-Fine: a Hierarchical Diffusion Model for Molecule Generation in 3D,'' in \textit{ICML}, 2023.
\bibitem{ref:seman2024} T. Jia, H. Li, C. Yang, T. Tao, and C. Shi, ``Graph Invariant Learning with Subgraph Co-mixup for Out-of-Distribution Generalization,'' in \textit{AAAI}, 2024.
\bibitem{ref:VAE} Martin Simonovsky and Nikos Komodakis, ``Graphvae: Towards generation of small graphs using variational autoencoders,'' in \textit{ICANN}, 2018.
\bibitem{ref:GAN} Nicola De Cao and Thomas Kipf, ``MolGAN: An implicit generative model for small molecular graphs,'' in \textit{ICML}, 2018.
\bibitem{ref:NF} Y. Luo, K. Yan, and S. Ji, ``Graphdf: A discrete flow model for molecular graph generation,'' in \textit{ICML}, 2021.
\bibitem{ref:Diff} H. Cao, C. Tan, Z. Gao, G. Chen, P. A. Heng, and S. Z. Li, ``A survey on generative diffusion model,'' \textit{arXiv preprint arXiv:2209.02646}, 2022.
\bibitem{ref:CF} F. Eijkelboom, G. Bartosh, C. A. Naesseth, M. Welling, and J. van de Meent, ``Variational Flow Matching for Graph Generation,'' in \textit{NeurIPS}, 2024.
\bibitem{ref:survey2020} A. Bonifati, I. Holubov\'a, A. P. P\'erez, and S. Sakr, ``Graph Generators: State of the Art and Open Challenges,'' in \textit{CSUR}, 2020.
\bibitem{ref:survey2022} Xiaojie Guo and Liang Zhao, ``A Systematic Survey on Deep Generative Models for Graph Generation,'' in \textit{TPAMI}, 2022.
\bibitem{ref:survey2023} C. Liu, W. Fan, Y. Liu, J. Li, H. Li, H. Liu, J. Tang, and Q. Li, ``Generative Diffusion Models on Graphs: Methods and Applications,'' in \textit{IJCAI}, 2023.
\bibitem{ref:RMAT} D. Chakrabarti, Y. Zhan, and C. Faloutsos, ``R-MAT: A recursive model for graph mining,'' in \textit{ICDM}, 2004.
\bibitem{ref:BTER} T. G. Kolda, A. Pinar, T. Plantenga, and C. Seshadhri, ``A scalable generative graph model with community structure,'' in \textit{SISC}, 2014.
\bibitem{ref:Darwini} S. Edunov, D. Logothetis, C. Wang, A. Ching, and M. Kabiljo, ``Darwini: Generating realistic large-scale social graphs,'' in \textit{arXiv preprint arXiv:1610.00664}, 2016.
\bibitem{ref:GCN} T. N. Kipf and M. Welling, ``Semi-supervised classification with graph convolutional networks,'' in \textit{ICLR}, 2017.
\bibitem{ref:CL} W. Aiello, F. Chung, and L. Lu, ``A Random Graph Model for Power Law Graphs,'' in \textit{Experimental Mathematics}, 2001.
\bibitem{ref:TUDataset} C. Morris, N. M. Kriege, F. Bause, K. Kersting, P. Mutzel, and M. Neumann, ``TUDataset: A collection of benchmark datasets for learning with graphs,'' \textit{arXiv preprint arXiv:2007.08663}, 2020.
\bibitem{ref:ER} Erd\H{o}s, P\'al and R\'enyi, Alfr\'ed, ``On Random Graphs I,'' in \textit{Publicationes Mathematicae}, 1959.
\bibitem{ref:BA} Albert, R. and Barab\'asi, L., ``Statistical mechanics of complex networks,'' in \textit{Reviews of Modern Physics}, 2002.
\bibitem{ref:WS} D. J. Watts and S. H. Strogatz, ``Collective dynamics of small-world networks,'' in \textit{Nature}, 1998.
\bibitem{ref:MMSB} E. M. Airoldi, D. M. Blei, S. E. Fienberg, and E. P. Xing, ``Mixed Membership Stochastic Blockmodels,'' in \textit{JMLR}, 2008.
\bibitem{ref:Kronecker} J. Leskovec, D. Chakrabarti, J. M. Kleinberg, C. Faloutsos, and Z. Ghahramani, ``Kronecker Graphs: An Approach to Modeling Networks,'' in \textit{JMLR}, 2010.
\bibitem{ref:GraphRNN} J. You, R. Ying, X. Ren, W. L. Hamilton, and J. Leskovec, ``GraphRNN: Generating Realistic Graphs with Deep Auto-regressive Models,'' in \textit{ICML}, 2018.
\bibitem{ref:GraphARM} L. Kong, J. Cui, H. Sun, Y. Zhuang, B. A. Prakash, and C. Zhang, ``Autoregressive Diffusion Model for Graph Generation,'' in \textit{ICML}, 2023.
\bibitem{ref:GraphILE} A. Bergmeister, K. Martinkus, N. Perraudin, and R. Wattenhofer, ``Efficient and Scalable Graph Generation through Iterative Local Expansion,'' in \textit{ICLR}, 2024.
\bibitem{ref:MMD} Gretton, A., Borgwardt, K. M., Rasch, M. J., Sch\"olkopf, B., and Smola, A., ``A kernel two-sample test,'' in \textit{JMLR}, 2012.
\end{thebibliography}
\end{document}


\appendix
\section{Notations and Definitions}
\begin{table}[!t]
    \centering
    \caption{Notations and Definitions.}
    \label{tab:note}
    \begin{tabular}{l|l}
        \hline
        \textbf{Symbols} & \textbf{Definition \& Description} \\
        \hline
        $\mathcal{G}$ & Graphs \\
        $\mathcal{V}, \mathcal{E}$ & Set of nodes and edges \\
        $N, M$ & Number of nodes and edges \\  
        $(N, M)$ & A graph with $N$ nodes and $M$ edges \\
        $\bar{D}, d_{max}$ & Average and maximal degree \\
        $d_u$ & Degree of node $u$ \\
        \hline
        $P(\lambda)$ & $\lambda$-Poisson Distribution \\
        $P(k_0|\lambda)$ & $\lambda$-Poisson Distribution when $X = k_0$ \\
        $k$ & Minimum positive integer satisfying $P(k|\lambda) < P(0|\lambda)$ \\
        \hline
        $\mathcal{G}_{sub}$ & Sequence of substructures \\
        $v_0^{(i)}$ & Anchor node of $\mathcal{G}_{sub}^{(i)}$ \\
        $D$ & Degree sequence of anchor nodes \\
        \hline
        $L_{ent}$ & Entry list, where $L_{ent}[i][j]$ is the degree of node $v_j^{(i)}$ \\
        $Pr(\cdot)$ & Node selection probability \\
        $s_{sub}^{(i)}$ & Node proportion of $\mathcal{G}_{sub}^{(i)}$ \\
        $P_E, P_e$ & Edge selection probability \\
        $c$ & Graph sparsity \\
        \hline
    \end{tabular}
\end{table}
Table~\ref{tab:note} shows the notations and definitions in the main text.

\section{Motivations}
In this section, we present a comprehensive rationale for the motivation behind our work.
First, we address the critical challenges posed by limited resources setting.
On one hand, we identify significant limitations in training data availability.
For example, users' relationships may have semantics~\cite{ref:seq2022} (e.g., motifs~\cite{ref:motif}).
In this case, conventional data anonymization techniques are not enough, as the implicit distribution patterns are leaked if the training data is available, causing severe security problems.
On the other hand, learning approaches may be limited.
For example, there are insufficient data and resources with quantity and quality assurance in some marginal application scenarios to support model learning.
It leads to the suffering of solutions such as deep generative models, large language models, etc., which require both data and training resources. 
Furthermore due to the black-box nature of the learning model, the learned invariant features do not have good interpretability.
Therefore, we need to discuss graph generation methods back to random graph theories in limited resources scenarios.

Next, we elaborate on the details of our method.
Since the resources are limited, there is an absence of prior knowledge regarding the true graph distribution, our major task is to capture robust invariant features that can accommodate a broader spectrum of graph categories, compensating for the scarcity of resources available.
Thereby, we draw inspiration from the scale-free properties observed in natural networks.
However, recognizing that strict reliance on a single characteristic may decay the generalization capability of generated graphs, we introduce a degree mixing strategy in the second stage of our framework.
This mechanism allows a weighted scaling based on the scale-free distribution while maintaining flexibility, which mitigates the rigid constraints imposed by single-feature dominance and optimizes the distribution of generated graphs through parameter adjustments.
Considering that Poisson distribution represents one of the most prevalent scale-free distributions, we leverage it as our primary sampling base.
We justify this selection in the experiment (see Sec $5.5^*$).
Note that we use $^*$ to represent the content in the main text.
We conduct comprehensive analyses with various distributions, including non-scale-free (e.g., Normal, Uniform) and approximate scale-free (e.g., Exponential, Pareto).
The experimental results demonstrate the superior performance of our method.

\section{Methodology}
\subsection{Algorithms}
\begin{algorithm}[!t]
    \caption{Build the Probability List}
    \label{alg:plist}
    \begin{algorithmic}[1] 
        \REQUIRE the entry list $L_{ent}$;
        \ENSURE the probability list $plist$, the indices list $nlist$;
        \STATE $plist \xleftarrow{} []$, $nlist \xleftarrow{} []$;
        \FOR{$i = 0, ..., l-1$} 
            \STATE compute $s_{sub}^{(i)}$ according to Equation $4^*$;
            \STATE $tlist \xleftarrow{} []$;
            \FOR{$j = 0, ..., N_{sub}^{(i)}$}
                \STATE compute $Pr(v_j^{(i)})$ according to Equation $3^*$;
                \STATE add index $[i, j]$ to $nlist$;
                \STATE add $Pr(v_j^{(i)})$ to $tlist$;
            \ENDFOR
            \STATE add $tlist$ to $plist$;
        \ENDFOR
        \STATE normalize $plist$ under Equation $6^*$;
        \STATE \textbf{return} $plist$, $nlist$
    \end{algorithmic}
\end{algorithm}
Algorithm~\ref{alg:plist} gives the procedure to build the probability list $plist$ through $L_{ent}$.
Since we sample anchor nodes based on the Poisson distribution, where $D \sim P(\bar{D})$, the expectation of the outer loop (line 2) is $\mathbb{E}(l) = \frac{N}{\bar{D} + 1}$, and the inner (line 5) is $\mathbb{E}(N_{sub}^{(i)}) = \bar{D} + 1$.
Thus, our method needs $\mathcal{O}(N)$ to build the probability list each time.
\begin{algorithm}[!t]
    \caption{Our Method: Hierarchical Scale-free Graph Generation}
    \label{alg:alg}
    \begin{algorithmic}[1] 
        \REQUIRE the number of nodes $N$, the number of edges $M$, the max degree $d_{max}$;
        \ENSURE the generated graph $\mathcal{G}$;
        \STATE $\bar{D} \xleftarrow{} \frac{2M}{N}$, $\mathcal{E} \xleftarrow{} \emptyset$;
        \STATE $\mathcal{G}_{sub} \xleftarrow{} [(d_{max} + 1, d_{max})]$, $sN \xleftarrow{} N - d_{max} - 1$;
        \WHILE{$sN > 0$}
            \STATE sample one possible $d \sim P(\bar{D})$; 
            \STATE add a star graph $(d + 1, d)$ to $\mathcal{G}_{sub}$;
            \STATE $sN \xleftarrow{} sN - d - 1$;
        \ENDWHILE
        \STATE build a degree list $L_{ent}$;
        \FORALL{$\mathcal{G}_{sub}^{(i)} \in \mathcal{G}_{sub}$}
            \STATE sample a node $u \in \mathcal{V}_{sub}^{(i)}$;
            \STATE build a probability list $plist$;
            \STATE sample a node $v \in \mathcal{V}_{sub}^{(j)}, (j \neq i)$ based on $plist$;
            \STATE add edge $e_{uv}$ to $\mathcal{E}$ and update $L_{ent}$;
        \ENDFOR
        \STATE build a graph $\mathcal{G}$ based on $\mathcal{G}_{sub}$ and $\mathcal{E}$;
        \STATE $\mathcal{E} \xleftarrow{} \emptyset$, $\mathcal{E}' \xleftarrow{} \emptyset$, $sM \xleftarrow{} M - |\mathcal{E}_\mathcal{G}|$;
        \WHILE{$sM > 0$}
            \STATE build a probability list $plist$;
            \STATE sample a pair $u, v$ based on $plist$;
            \STATE add edge $e_{uv}$ to $\mathcal{E}$ and update $L_{ent}$;
            \STATE $sM \xleftarrow{} sM - |\mathcal{E}|$, $\mathcal{E}' \xleftarrow{} \mathcal{E}' \cup \mathcal{E}$, $\mathcal{E} \xleftarrow{} \emptyset$;
        \ENDWHILE
        \STATE add edges in $\mathcal{E}'$ to $\mathcal{G}$;
        \STATE \textbf{return} $\mathcal{G}$
    \end{algorithmic}
\end{algorithm}
The overall framework is in Algorithm~\ref{alg:alg}.
Our method contains three parts.
Firstly, we select anchor nodes and build the initial star graphs in the first stage (lines 2-7).
Next, we ensure the previous graph connection at the beginning of the second stage (lines 9-14).
And finally, we create the remaining edges (lines 17-22).
Note that our method is hierarchical, and each part includes an update of the probability, which guarantees the accuracy of the generation.
\subsection{Theoretical Proof}
Our method partitions $\mathcal{V}$ into two sets of nodes: a set $\mathcal{V}_{a}$ of anchors and a set $\mathcal{V}_{na}$ of non-anchors.
$D \sim P(\bar{D})$ initializes $\mathcal{V}_a = \{v_0^{(0)}, ..., v_0^{(l-1)}\}$, thus $\mathcal{V}_a \sim P(\bar{D})$.
$\mathcal{V}_{na}$ calculates the sampling probability according to Equation $3^*$, where we have
\begin{equation}
    \begin{aligned}
        \mathbb{E}(\mathcal{V}_{na}) & = \sum_d d \cdot \bar{Pr}(v_{na}; d) \\
        & = \sum_d d \cdot \left[\sum_i s_{sub}^{(i)} \cdot P(d|\bar{D})\right] \\
        & = \sum_d d \cdot \left[P(d|\bar{D}) \cdot \sum_i s_{sub}^{(i)}\right] \\
        & = \left[\sum_d d \cdot P(d|\bar{D})\right] \cdot \underset{1}{\underbrace{\left[\sum_i s_{sub}^{(i)}\right]}} \\
        & = \mathbb{E}\left[d \sim P(\bar{D})\right] \cdot 1 = \bar{D}. \\
    \end{aligned}
    \label{eq:na}
\end{equation}
In the actual generation process, there are a small number of nodes $\mathcal{V}_{\epsilon}$ out of distribution due to the scaling of Equation $4^*$ and the hierarchical sampling strategy in Figure $3^*$.
However, since $|\mathcal{V}_{\epsilon}| \ll |\mathcal{V}|$, our method maintains the conclusion in Observation $1^*$.

We further discuss the expectation of Equation $3^*$.
Considering that anchor nodes guide the computation of Equation $4^*$, $s_{sub}^{(i)}$ is independent of the non-anchor node degree selection.
So we have:
\begin{equation}
    \begin{aligned}
        \mathbb{E}\left[Pr(v_j^{(i)})\right] & = \mathbb{E}\left[s_{sub}^{(i)} \cdot P(L_{ent}[i][j] + 1|\bar{D})\right] \\
        & = \underset{Eq. 4^*}{\underbrace{\mathbb{E}\left[s_{sub}^{(i)}\right]}} \cdot \mathbb{E}\left[P(L_{ent}[i][j] + 1|\bar{D})\right] \\
        & = \frac{\mathbb{E}\left[N_{sub}^{(i)}\right]}{N} \cdot P(\overset{Eq.~\ref{eq:na}}{\overbrace{\mathbb{E}\left(L_{ent}[i][j] + 1\right)}}|\bar{D}) \\
        & = \frac{\bar{D} + 1}{N} \cdot P(\bar{D} | \bar{D}) = \frac{e^{-\bar{D}}(\bar{D} + 1)\bar{D}^{\bar{D}}}{N\bar{D}!}. \\
    \end{aligned}
    \label{eq:node}
\end{equation}

It means that we no longer consider equivalently in node selection.
Given the sparsity $c$, our method can maintain a probability expectation higher than $\frac{1}{N}$ for a large number of nodes.
For example, when $c=0.2358$, Equation~\ref{eq:node} exceeds the average selection if $N \geq 20$.

In addition, according to~\cite{ref:ER}, we have the following theorem:
\begin{theorem}[Connectivity]
    To ensure that the graph is connected, the number of edges should satisfy
    \begin{equation}
        M = [\frac{1}{2}N\log N + \epsilon N],
        \label{eq:M}
    \end{equation}
    where $[x]$ is the integer part of $x$ and $\epsilon$ is an arbitrary fixed real number.
    \label{th:ER}
\end{theorem}

We can obtain three thresholds concerning the edge probability $P_E$ of random graphs, proved by~\cite{ref:ER,ref:CSW}.
The first is $P_2 = \frac{\bar{D}}{N-1}$, a plain selection for $\bar{D}$ edges drawing from a center node.
The second is $P_3 = \frac{\log N}{N - 1}$, showing fewer isolated nodes.
The last is $P_4 = \frac{2\log N}{N - 1}$, ensuring no isolated nodes exist. 
This series of probabilities is essentially a consideration of the sparsity of the graph, where the expected number of edges is $M = \frac{N(N-1)}{2} \cdot P_E$, drawing that $c = P_E$.
Taking $P_4$ as an example, our method picks one node with a higher probability than the average selection when $N \geq 142$.
\subsection{Time Complexity}
The time complexity of our proposed method contains three parts.
The first part is the parse phase.
Given a graph $(N, M)$, our method generates initial star graphs under the Poisson distribution of $\mathcal{P}(\lambda)$, with $\mathcal{O}(N + \frac{N}{\lambda})$ time complexity.
The second part connects sub-structures, scanning each supernode once.
For possible edge connecting $\mathcal{G}_{sub}^{(1)}$ and $\mathcal{G}_{sub}^{(2)}$ picked, our method decides which entries $u$ and $v$ the edge belongs to, where $u \in \mathcal{V}_{sub}^{(1)}, v \in \mathcal{V}_{sub}^{(2)}$, which costs $\mathcal{O}\left(\frac{N}{\lambda}\cdot2\lambda + \frac{3N}{\lambda}\right)$.
The third part is to rebuild the rest of the $M - N$ edges, which costs $\mathcal{O}(\mathcal{K}N + \sum_{\mathcal{K}}\Bar{M}\cdot\mathcal{O}(1))$ in total, where $\mathcal{K}$ is the number of scans and $\Bar{M}$ is the average edges picked each time.
Our method generates average $\frac{M}{\mathcal{K}}$ edges and updates the probability matrix like the second part.
We could estimate $\mathcal{K} \simeq \log M$ in a logarithmic descent speed.
Thus, the time complexity of the third part is $\mathcal{O}(N\log M + M)$.
The final time complexity is the sum of all three parts above.
To guarantee the graph connectivity, we approximate $M$ with the probability of $P_E = \frac{\log N}{N-1}$, where $M = \frac{N(N-1)}{2} \times P_E = \frac{N\log N}{2}$.
Since $\lambda = \bar{D}$, the time complexity above is equal to $\mathcal{O}[\frac{4N}{\log N} + 3N + N\log(\frac{N\log N}{2}) + \frac{N\log N}{2}]$.

In practice, however, edges are far more than nodes, where $M = c\times\frac{N(N-1)}{2}, c\in(0, 1]$.
It shows that $M \propto N^2$ rather than $N\log N$.
So the complexity eventually bounds to $\mathcal{O}(N^2)$ (constants omitted).
It is similar to the plain method of comparing all node pairs.
\subsection{Limitation}
\begin{table}[!t]
    \centering
    \caption{Advanced details of all 12 datasets.}
    \begin{tabular}{llrrrr}
        \hline
         Dataset & Category & deg. & clus. & orbit & sparse \\
         \hline
         ENZYMES & bioinformatics & 3.86 & 0.45 & 3.92 & 0.16 \\
         MUTAG & molecule & 2.19 & 0.00 & 1.14 & 0.14 \\
         NCI1 & molecule & 2.16 & 0.00 & 1.11 & 0.09 \\
         PROTEINS & bioinformatics & 3.78 & 0.51 & 3.29 & 0.21 \\
         deezer\_ego\_nets & social & 4.29 & 0.51 & 54.9 & 0.23 \\
         IMDB-BINARY & social & 8.89 & 0.95 & 40.3 & 0.52 \\
         IMDB-MULTI & social & 8.10 & 0.97 & 27.2 & 0.77 \\
         REDDIT-BINARY & social & 2.34 & 0.05 & 4.2k & 0.02 \\
         CLUS & synthetic & 5.91 & 0.66 & 18.4 & 0.42 \\
         EGO & synthetic & 1.99 & 0.00 & 6.32 & 0.01 \\
         GRID & synthetic & 3.11 & 0.00 & 3.11 & 0.12 \\
         TREE & synthetic & 1.86 & 0.00 & 0.83 & 0.14 \\
         \hline
    \end{tabular}
    \label{tab:datasets}
\end{table}
Since our method relies on Observation $1^*$, our performance degrades for partial graphs without the scale-free property.
Experimental results further illustrate this limitation (Sec. V-B in the main text).
The closer the ground truth distribution is to a strict power law, the better our generation is.

\section{Experimental Settings}
\subsection{Datasets}
Datasets are from three categories: bioinformatics \& molecule (a-d), social networks (e-h), and synthetic (i-l).
Descriptions are listed below.

\textbf{(a) ENZYMES} is derived from the BRENDA enzyme database, which shows the protein tertiary structures. It contains 600 graphs from 6 enzyme classes, each having average nodes and edges of 32.6 and 62.1.
\textbf{(b) MUTAG} represents chemical compounds, having 188 samples to predict their mutagenicity on Salmonella typhimurium. It is a smaller dataset with average nodes and edges of 17.9 and 19.8.
\textbf{(c) NCI1} is a dataset of chemical compounds the same as the MUTAG. Among 4,110 graphs, each has an average of 29.9 nodes and 32.3 edges, categorized by positive or negative against cell lung cancer.
\textbf{(d) PROTEINS} is a dataset of proteins classified by whether they are enzymes or not. There are 1,113 graphs with an average of 39.1 nodes and 72.8 edges.

\textbf{(e) deezer\_ego\_nets} is a user friendship network from the music service Deezer. It has 9,629 graphs, each having an average of 23.5 nodes and 65.3 edges.
\textbf{(f) IMDB-BINARY} shows a movie collaboration of 1,000 actors/actresses in IMDB. These graphs come from two categories: Action and Romance. Each graph has 19.8 nodes and 96.5 edges.
\textbf{(g) IMDB-MULTI} is the same as its BINARY version, except for the multi-class graphs. It has 1,500 graphs, with average nodes and edges of 13.0 and 65.9.
\textbf{(h) REDDIT-BINARY} consists of 2,000 graphs of online discussions on Reddit, labeled by QA-based or discussion-based community. It is our largest dataset, each having 429.6 nodes and 497.8 edges on average.

All datasets above are real-world and from TUDataset~\cite{ref:TUDataset}.
Next are four synthetic datasets drawing various characteristics, each containing 200 graphs.
\textbf{(i) CLUS} is the graph from the Holme and Kim algorithm~\cite{ref:CLUS}, with power-law degree distribution and approximate average clustering.
\textbf{(j) EGO} generates ego networks from power law degree sequences with average nodes of 150. 
\textbf{(k) GRID} is standard 2D grid graph from 10x10 to 20x20.
\textbf{(l) TREE} generates a tree with 100 to 200 nodes converted from a uniformly random Pr\"{u}fer sequence.

Furthermore, Table~\ref{tab:datasets} shows four more advanced details: average degree, clustering coefficient, orbit count, and sparsity.
\subsection{Implementation Details}
\begin{table}[!t]
    \centering
    \caption{Parameter Settings.}
    \begin{tabular}{c|cl}
         \hline
         \textbf{Methods} & $f(x)$ & Default Settings \\
         \hline
         Uniform & $1 / (b-a)$ & equal probability \\
         Normal & $e^{-(x-\mu)^2 / 2\sigma^2} / \sqrt{2\pi}\sigma$ & $\mu = \bar{D}, \sigma = 1$ \\
         Exponential & $\lambda e^{-\lambda x}$ & $\lambda = \bar{D}$ \\
         Gamma & $\beta^\alpha x^{\alpha-1}e^{-\beta x} / \Gamma(\alpha)$ & $\alpha = \bar{D}, \beta = P_E$ \\
         Pareto & $kx^k_{min} / x^{k+1}$ & $k = \bar{D}, x_{min} = d_{min}$ \\
         Ours & $\lambda^x e^{-\lambda} / x!$ & $\lambda = \bar{D}$ \\
         \hline
    \end{tabular}
    \label{tab:deg_setting}
\end{table}
We use the same sources as~\cite{ref:GraphRNN} to conduct four traditional methods, while WS is from the open toolkit NetworkX~\cite{ref:nx}.
As for deep learning SOTAs, we try our best to reproduce them based on their open-source codes.
All experiments are trained and evaluated on an NVIDIA RTX 3050 OEM 24GB GPU and 32GB memory.

\section{Experiment Results}
\subsection{Parameter Settings}
\begin{table*}[!t]
    \centering
    \caption{MMD evaluation between deep learning SOTAs and our method.}
    \begin{tabular}{c|ccc|ccc|ccc|ccc}
        \hline
        \multirow{2}{*}{} & \multicolumn{3}{c}{ENZYMES} & \multicolumn{3}{c}{deezer\_ego\_nets} & \multicolumn{3}{c}{CLUS} \\
        Methods & deg. & clus. & orbit & deg. & clus. & orbit & deg. & clus. & orbit \\
        \hline
        GraphRNN & 1.6851 & 1.0127 & 1.0231 & 1.2262 & 1.3054 & 0.5834 & 0.2489 & 0.7418 & 0.8339 \\
        +training & \underline{0.2042} & 1.2339 & \textbf{0.1088} & 1.1688 & 1.3245 & 1.1597 & 0.5203 & 0.7172 & 0.6988 \\
        \hline
        GraphARM & 1.8721 & 1.7257 & 0.4734 & 1.3041 & 1.2882 & 1.0891 & 1.3535 & 1.4196 & 1.1840 \\
        +training & 1.7830 & 1.3862 & 0.2326 & 0.8506 & 0.9830 & 1.2827 & 1.3362 & 1.4200 & 0.7786 \\
        \hline
        GraphILE & 0.7302 & 1.2613 & 0.7600 & 0.6224 & 1.1985 & 1.2144 & \underline{0.1293} & \underline{0.4000} & 0.4538 \\
        +training & 0.2333 & \textbf{0.2603} & \underline{0.1121} & \underline{0.2295} & \underline{0.4003} & \underline{0.5025} & 0.2118 & 0.9729 & \underline{0.2886} \\
        \hline
        Ours & \textbf{0.1811} & \underline{0.4621} & 0.1271 & \textbf{0.0371} & \textbf{0.0668} & \textbf{0.0026} & \textbf{0.0840} & \textbf{0.1911} & \textbf{0.0509} \\
        \hline
        \hline
        \multirow{2}{*}{} & \multicolumn{3}{c}{MUTAG} & \multicolumn{3}{c}{IMDB-BINARY} & \multicolumn{3}{c}{EGO} \\
        Methods & deg. & clus. & orbit & deg. & clus. & orbit & deg. & clus. & orbit \\
        \hline
        GraphRNN & 1.6539 & 1.9100 & 1.2918 & 1.0155 & 1.8265 & \underline{0.4083} & 1.8982 & 1.6031 & 1.0489 \\
        +training & \underline{0.2138} & 0.7723 & \underline{0.0456} & \underline{0.5036} & 1.8412 & 0.7765 & 0.9116 & 0.4883 & 0.4290 \\
        \hline
        GraphARM & 1.7245 & 0.0000 & 0.0549 & 1.3703 & 1.8764 & 1.3149 & 1.4053 & 0.0000 & 0.0303 \\
        +training & 1.6387 & \textbf{0.0000} & 0.0532 & 1.1919 & 1.3964 & 1.2144 & 1.4160 & 0.0000 & \underline{0.0286} \\
        \hline
        GraphILE & 0.5061 & 1.2001 & 1.0278 & 0.6126 & \underline{1.3221} & 1.2408 & 0.6892 & 0.0000 & 0.5790 \\
        +training & 0.5833 & 1.0054 & 0.9027 & 0.5981 & 1.3465 & 0.4415 & \underline{0.6708} & 0.0000 & 0.5540 \\
        \hline
        Ours & \textbf{0.0355} & \underline{0.0561} & \textbf{0.0014} & \textbf{0.0393} & \textbf{0.7161} & \textbf{0.1436} & \textbf{0.0382} & \textbf{0.0000} & \textbf{0.0080} \\
        \hline
        \hline
        \multirow{2}{*}{} & \multicolumn{3}{c}{NCI1} & \multicolumn{3}{c}{IMDB-MULTI} & \multicolumn{3}{c}{GRID} \\
        Methods & deg. & clus. & orbit & deg. & clus. & orbit & deg. & clus. & orbit \\
        \hline
        GraphRNN & 1.7020 & 1.9439 & 1.1536 & 1.2608 & 1.7468 & 1.1993 & 1.8623 & 1.9928 & 1.0156 \\
        +training & 0.3920 & 0.6467 & 0.0825 & 0.6925 & 1.5325 & 0.8245 & 0.8239 & \underline{0.1763} & 1.1837 \\
        \hline
        GraphARM & 1.7009 & 0.0371 & 0.0470 & 0.7155 & 0.8996 & - & 1.9733 & 0.0000 & 0.4525 \\
        +training & 1.6999 & \textbf{0.0000} & \underline{0.0464} & 0.9056 & \textbf{0.0304} & - & 1.9460 & \textbf{0.0000} & \underline{0.3720} \\
        \hline
        GraphILE & 0.6607 & 1.2002 & 1.1948 & 0.3850 & 1.0126 & 0.4014 & 0.6400 & 1.2003 & 1.2129 \\
        +training & \underline{0.0446} & 0.7988 & 0.1271 & \underline{0.3742} & 0.8134 & \underline{0.3998} & \underline{0.4095} & 2.0000 & 1.0415 \\
        \hline
        Ours & \textbf{0.0340} & \underline{0.0263} & \textbf{0.0002} & \textbf{0.0160} & \underline{0.3882} & \textbf{0.0416} & \textbf{0.2459} & 0.4081 & \textbf{0.0148} \\
        \hline
        \hline
        \multirow{2}{*}{} & \multicolumn{3}{c}{PROTEINS} & \multicolumn{3}{c}{REDDIT-BINARY} & \multicolumn{3}{c}{TREE} \\
        Methods & deg. & clus. & orbit & deg. & clus. & orbit & deg. & clus. & orbit \\
        \hline
        GraphRNN & 1.6673 & 1.0854 & 0.8910 & 1.8544 & \underline{0.7590} & \underline{0.9275} & 1.4897 & 1.9996 & 1.1506 \\
        +training & 0.2860 & 0.6573 & 0.3239 & \underline{0.6891} & \textbf{0.1295} & 1.1999 & 0.7043 & 0.0892 & 0.3258 \\
        \hline
        GraphARM & 1.9206 & 1.3667 & 0.2019 & - & - & - & 1.5013 & 0.0000 & 0.0233 \\
        +training & 1.7909 & 1.3593 & 0.3079 & - & - & - & 1.4871 & 0.0000 & \underline{0.0216} \\
        \hline
        GraphILE & 0.6023 & 0.5941 & 0.4550 & - & - & - & 0.6348 & 1.2000 & 0.0243 \\
        +training & \underline{0.2730} & \underline{0.5773} & \underline{0.1888} & - & - & - & \underline{0.5789} & 1.0051 & 0.9097 \\
        \hline
        Ours & \textbf{0.2728} & \textbf{0.5245} & \textbf{0.1635} & \textbf{0.3229} & 1.3575 & \textbf{0.0231} & \textbf{0.0013} & \textbf{0.0000} & \textbf{0.0003} \\
        \hline
    \end{tabular}
    \label{tab:sota}
\end{table*}
Table~\ref{tab:deg_setting} shows the detailed parameter settings.
\subsection{Comparison with Deep Generative Models}
Although deep generative models critically suffer when no training data is available, we are still interested in their performance.
Table~\ref{tab:sota} shows the results.
For each baseline, we first evaluate all datasets with only the initialization state, which simulates the situation of no training data.
Next, we train all models with a fixed number of steps, assuming there are limited numbers for fetching training data.
Results show that the limitation of training data has disastrous effects on deep learning models.
On the one hand, models cannot work competently without training.
On the other hand, the undertrained model does not have enough time to learn implicit characteristics in the dataset, leading to the instability of early results.
In most cases, models fail to optimize the results of all three matrices simultaneously.
However, the training always works.
Models outperform our method after limited training steps under partial datasets.
To sum up, our method retains its advantage when training is impossible.
In the case of limited training, our method can achieve stable and appreciable results under a fixed training step.
\subsection{Comparison with Large Language Models}
\begin{table*}[!t]
    \centering
    \caption{MMD evaluation between LLM and our method.}
    \begin{tabular}{c|ccc|ccc|ccc}
        \hline
        \multirow{2}{*}{} & \multicolumn{3}{c}{ENZYMES} & \multicolumn{3}{c}{deezer\_ego\_nets} & \multicolumn{3}{c}{CLUS} \\
        Methods & deg. & clus. & orbit & deg. & clus. & orbit & deg. & clus. & orbit \\
        \hline
        LLM & 0.3611 & 1.6081 & 0.9360 & 0.3171 & 0.7379 & 0.3207 & 0.2709 & 0.4550 & 0.2787 \\
        Ours & 0.3274 & 1.3823 & 0.6577 & 0.1401 & 0.3127 & 0.0686 & 0.1437 & 0.2357 & 0.0998 \\
        \hline
        \hline
        \multirow{2}{*}{} & \multicolumn{3}{c}{MUTAG} & \multicolumn{3}{c}{IMDB-BINARY} & \multicolumn{3}{c}{EGO} \\
        Methods & deg. & clus. & orbit & deg. & clus. & orbit & deg. & clus. & orbit \\
        \hline
        LLM & 0.0437 & 0.1453 & 0.0029 & 0.2373 & 0.7349 & 0.2542 & 0.3846 & 0.0055 & 0.4626 \\
        Ours & 0.0346 & 0.0961 & 0.0011 & 0.1143 & 0.7154 & 0.2242 & 0.2774 & 0.0055 & 0.0938 \\
        \hline
        \hline
        \multirow{2}{*}{} & \multicolumn{3}{c}{NCI1} & \multicolumn{3}{c}{IMDB-MULTI} & \multicolumn{3}{c}{GRID} \\
        Methods & deg. & clus. & orbit & deg. & clus. & orbit & deg. & clus. & orbit \\
        \hline
        LLM & 0.0551 & 0.2086 & 0.0072 & 0.1118 & 0.5709 & 0.1303 & 0.0845 & 0.8327 & 0.0138 \\
        Ours & 0.0114 & 0.0219 & 0.0004 & 0.0762 & 0.5103 & 0.1039 & 0.1697 & 0.5414 & 0.0080 \\
        \hline
        \hline
        \multirow{2}{*}{} & \multicolumn{3}{c}{PROTEINS} & \multicolumn{3}{c}{REDDIT-BINARY} & \multicolumn{3}{c}{TREE} \\
        Methods & deg. & clus. & orbit & deg. & clus. & orbit & deg. & clus. & orbit \\
        \hline
        LLM & 0.3283 & 0.6020 & 0.2683 & 1.1578 & 0.1051 & 1.0935 & 0.0269 & 0.1864 & 0.0007 \\
        Ours & 0.2625 & 0.5823 & 0.2194 & 0.4023 & 1.5153 & 0.1999 & 0.0057 & 0.0000 & 0.0007 \\
        \hline
    \end{tabular}
    \label{tab:llm}
\end{table*}
\begin{table}[!t]
    \centering
    \caption{Detailed characteristics of large datasets.}
    \begin{tabular}{lcccrrrr}
        \hline
        Dataset & $|\mathcal{G}|$ & $\bar{N}$ & $\bar{M}$ & deg. & clus. & orbit & sparse \\
        \hline
        PPI & 20 & 2,245.3 & 61,318.4 & 25.8 & 0.18 & 2.5E4 & 1.2E-2 \\
        Flickr & 1 & 89,250.0 & 899,756.0 & 10.1 & 0.03 & 1.1E5 & 1.1E-4 \\
        \hline
    \end{tabular}
    \label{tab:large}
\end{table}
In recent years, there has been a growing trend of exploring the utilization of Large Language Models (LLMs) or agents for generating scale-free networks.
Within our limited resources problem setting, training a LLM or agent from scratch using extensive training data is impossible.
However, considering that the powerful tools such as ChatGPT and DeepSeek are available and convenient to integrate, we conduct some experiments between our method and LLM under the neglect of potential data leaking risks.
We formalize the graph generation task and establish standardized I/O formats, which are treated as prompt for LLM processing, as follows:
\begin{itemize}
    \item Now you have a graph generation task in a limited resources scenario where you don't know the distribution of the real graphs.
    \item Please just use the number of nodes, edges and the maximum node degree to generate graphs that matches the real ones as closely as possible.
    \item The input is multiple triples, each of the format: $(N, M, d)$, where $N$ and $M$ are the number of nodes and edges, $d$ is the maximum node degree.
    \item Output a list for each triple in the form of edge list.
\end{itemize}
We conduct experiments on both ChatGPT and DeepSeek and select the best results to report.
The results are in Table~\ref{tab:llm}.
Despite LLMs' extensive generalization capabilities, they demonstrate insufficient competence in generating graphs within limited resources problem setting.
We also observe some interesting phenomenons.
When the number of nodes and edges are relatively small, LLMs generate edge lists directly.
Meanwhile, when the number increases, LLMs tend to provide a code piece for generation.
Both ChatGPT and DeepSeek propose code implementations similar to EA, generating graphs by selecting edges randomly while not exceeding the maximum degree constraint.
As we give LLMs multiple graph generation alternatives, LLMs show a pronounced preference for BA.
\subsection{Time Analysis}
\begin{table*}[!t]
    \centering
    \caption{Ruintime evaluation between baselines and our method.}
    \begin{tabular}{c|cccr|cccr}
        \hline
        \multirow{2}{*}{} & \multicolumn{4}{c}{PPI} & \multicolumn{4}{c}{Flickr} \\
        Methods & deg. & clus. & orbit & time & deg. & clus. & orbit & time \\
        \hline
        ER & 0.3613 & 1.4957 & 0.1000 & 1.44E-01 & 2.0000 & 0.0965 & 2.0000 & 1.75E+02 \\
        BA & 0.2683 & 1.0549 & 0.1000 & 5.37E-02 & 2.0000 & 0.0911 & 2.0000 & 1.20E+00 \\        
        WS & 0.3397 & 1.9481 & 0.1167 & 1.86E-02 & 2.0000 & 2.0000 & 2.0000 & 4.35E-01 \\
        MMSB & - & - & - & - & - & - & - & - \\
        Kronecker & - & - & - & - & - & - & - & - \\
        \hline
        Ours & 0.2224 & 0.1598 & 0.1000 & 1.09E-01 & 2.0000 & 0.0861 & 2.0000 & 1.70E+00 \\
        \hline
    \end{tabular}
    \label{tab:time}
\end{table*}
\begin{figure}[!t]
    \centering
    \includegraphics[width=\linewidth]{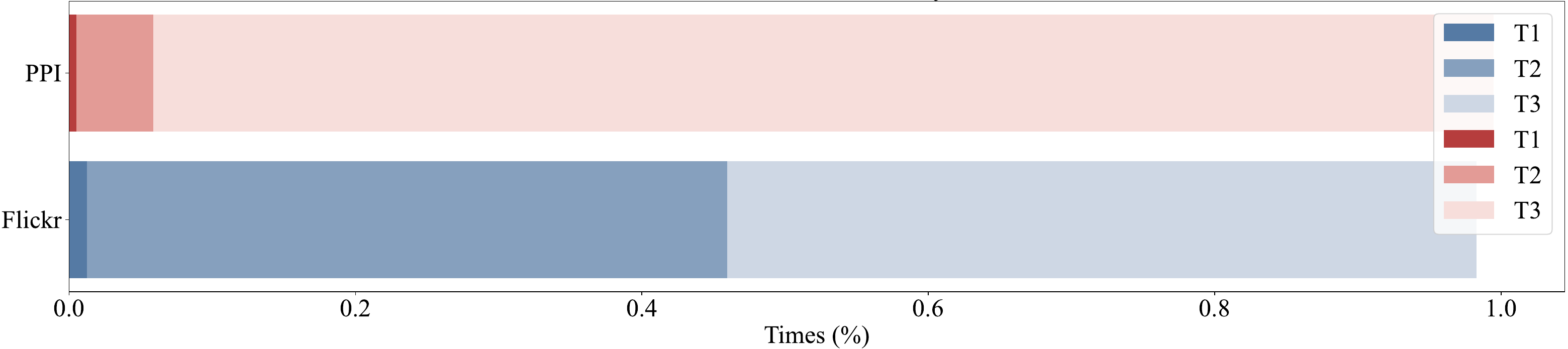}
    \caption{Detailed Runtime Analysis. T1: the parse phase. T2: connecting sub-structures. T3: rebuilding the rest edges.}
    \label{fig:large}
\end{figure}
We introduce several large datasets to make a further evaluation of time complexity of our method.
The detailed characteristics of datasets are in Table~\ref{tab:large}.
The experimental results, as presented in Table~\ref{tab:time}, demonstrate that our proposed method achieves an optimal balance between performance quality and computational efficiency.
In terms of graph generation quality, we outperform all comparative approaches.
In terms of computational efficiency, our method exhibits superior time complexity, with an average runtime second only to BA.
The theoretical worst-case time complexity of our method is $\mathcal{O}(N^2)$, which aligns with the complexity of EA.
It is noteworthy that WS has a shorter generation time, but its inherent limitation of initializing graphs with ring structures significantly compromises the quality of graphs generated, rendering it less effective for practical scenarios.
MMSB and Kronecker are fail to generate graphs due to out of time or memory.
Furthermore, we provide a detailed time analysis, decomposing the runtime according to the generation phase, as detailed in Figure~\ref{fig:large}.
The proportional relationship among T1, T2, and T3 aligns closely with the analysis above in Section C.3, where T3 accounts for the majority of computational resources.